\documentclass[usenatbib]{mn2e}
\loadboldmathitalic 
\title []{An H$\alpha$ survey of the rich cluster A\,1689}

\author[Balogh \etal]
{Michael L. Balogh$^{1,4}$, Warrick J. Couch$^{2}$, Ian Smail$^{1}$, Richard G. Bower$^{1}$  \newauthor
\& Karl Glazebrook$^{3}$\\
$^{1}$Department of Physics, University of Durham, South Road, Durham, DH1 3LE, UK\\
$^{2}$School of Physics, University of New South Wales, Sydney 2052, Australia\\
$^{3}$Department of Astrophysics, Johns Hopkins University, Baltimore, Maryland, USA\\
$^{4}$email:M.L.Balogh@durham.ac.uk\\
}
\date{\today}
\def\lesssim{\mathrel{\hbox{\rlap{\hbox{\lower4pt\hbox{$\sim$}}}\hbox{$<$}}}}
\def\gtrsim{\mathrel{\hbox{\rlap{\hbox{\lower4pt\hbox{$\sim$}}}\hbox{$>$}}}}
\def\etal{{\it et al.\thinspace}}

\def\ha{H$\alpha$}
\def\6m{6.7$\micron$}
\def\15m{15$\micron$}
\def\oii{[O{\sc ii}]}
\def\nii{[N{\sc ii}]}

\usepackage{epsfig}
\usepackage{amsmath}
\begin{document} 
\maketitle 
\begin{abstract}
We present results of an H$\alpha$ survey in the rich cluster A\,1689 at $z=0.18$,
using the LDSS++ spectrograph on the AAT.  We obtained spectra covering redshifted
H$\alpha$ at $z=0.16$--$0.22$, 
for 522 galaxies brighter than $I=22.5$, covering a field of 8.7\arcmin$\times$8.7\arcmin
($\sim 1.1\times1.1 h^{-1}$ Mpc at $z=0.18$).
We securely detect H$\alpha$ emission in 46 of these galaxies; accounting for selection effects
due to sampling and cluster membership, we determine that $24\pm4$\% of cluster members 
brighter than $M_R=-16.5+5\log{h}$ are
detected with H$\alpha$ flux greater than $4h^{-2}\times10^{38}$ ergs s$^{-1}$.  This corresponds
to a limiting star formation rate of 0.008 $h^{-2} M_\odot$ yr$^{-1}$, assuming 1 magnitude of dust extinction.
Our observations are sufficiently sensitive to 
detect galaxies with star formation rates comparable to that of the Milky Way ($\gtrsim 3 M_\odot$ yr$^{-1}$), unless they
are obscured by more than 7 magnitudes of extinction.  From a {\it Hubble Space Telescope}
mosaic covering $7.5'\times10.0'$, we determine morphologies for 199 galaxies brighter than $I=21$, and find that $\sim$20\% of
the cluster members are of type Sa or later.  More than $90$\% of
cluster spirals show H$\alpha$ emission, compared with less than $10$\% of E and S0 galaxies.  The cluster H$\alpha$ luminosity
function has a low normalisation relative to the $z\sim0.2$ field, by $\sim 50\%$, after accounting
for the different fraction of spiral galaxies in the two environments.  
When compared with local field galaxies, this suggests that star formation activity is suppressed in
early-type cluster galaxies, relative to their field counterparts.  Our sample includes 29 galaxies
previously observed with {\it ISOCAM} at \6m\ and \15m.  
We detect all \15m\ sources at H$\alpha$, so there is no evidence for any star
formation completely hidden at H$\alpha$.  Comparing the \15m\ and H$\alpha$ fluxes, we find
evidence that some mid-infrared-detected galaxies could be obscured by as much as 3 magnitudes of
extinction at $H\alpha$, although this depends on the largely
unknown contribution from any AGN-heated dust to the mid-infrared flux.  
\end{abstract}
\section{Introduction}
There is an increasing body of observational evidence
that star formation in the cores of galaxy clusters is much lower than
that in the surrounding field, whatever the redshift \citep[e.g.,][hereafter Paper~I]{B+97,B+98,P+99,Martin00,C+01}.
Although there is evidence that higher redshift clusters have a larger
population of late-type, current or recently star-forming galaxies than local clusters
\citep[e.g.,][]{BO84,C+98,D+97,Margo,KB01}, a similar increase in
activity is also seen in the field \citep{CFRS6,Madau,Cowie+99}.  Thus, the increase
in cluster activity with redshift could simply reflect this, modulated by  the changing
rate of infall onto the clusters \citep{PSext,Erica}.

However, most of the spectroscopic evidence for current or recent star formation in moderate to
high redshift clusters comes from observations of the \oii$\lambda$3727 emission line.  Unfortunately, 
\oii\ emission is sensitive to both metallicity effects and dust
extinction \citep{Kenn_review,Jansen}, and in both these respects the H$\alpha$ emission line provides a 
superior indicator of star formation.  At moderate redshifts, however, H$\alpha$ is redshifted
into a wavelength regime dominated by bright sky lines, and its observation requires
selecting a cluster for which H$\alpha$ lies in a window between these lines \citep[Paper~I;][]{A2390_BM}.

There remains the intriguing possibility that
a substantial amount of star formation (in clusters and elsewhere) is
obscured by dust, thus rendering even H$\alpha$ surveys incomplete 
\citep{Blain99,Smail-radio,DO99,PW00}.   Dust heated by star formation activity
will radiate at mid-infrared (MIR) wavelengths, and thus such optically-obscured
star formation may be recovered in MIR surveys.
Evidence suggests that most
of the MIR emission in local spiral galaxies is due to star formation 
\citep{Genzel98,Lutz98,Fadda02,RSVB}, but this is still controversial because
the spectral shape of any contribution to the MIR from AGN-heated dust
is uncertain \citep{Tran01}.
If most of the MIR emission originates from starburst-heated
dust, then observations suggest that much of the star formation in the universe
is optically obscured by this dust \citep{RR+97,Flores99,RE99,ISO-HDF}.  
Further complicating the issue is that the MIR population evolves strongly
with redshift, which could indicate that the relative contributions of AGN-
and starburst-heated dust are strongly redshift dependent \citep{ALB,FI,A+01,WFG}.  
Such complications make the analysis and interpretation of current MIR surveys time-consuming.

Recently, \citet{Fadda} presented {\it ISOCAM} observations of the $z=0.18$
cluster A\,1689 at \6m\ and \15m.  At the cluster redshift, 
the \6m\ filter covers the tail of stellar emission and aromatic
carbon compounds associated with star formation, while the \15m\ filter is 
dominated by the hot dust continuum.
\citet{Fadda} showed that A\,1689 contains an excess of
\15m\ sources relative to local clusters, suggesting that the cluster
hosts a population of dusty starburst galaxies.  
We observed A\,1689 in April 2001,
with the upgraded Low Dispersion Survey Spectrograph \citep[LDSS++,][]{GBH} on the 3.9m
Anglo-Australian Telescope (AAT), as part of a larger program to measure H$\alpha$ in a diverse sample
of four $z\sim 0.2-0.3$ clusters (see Paper~I for initial results). 
Our observational technique allows deep, efficient measurements of H$\alpha$ to
be made, and moreover gives precise dynamical information and allows us to identify 
the presence of AGN from the relative
strength of the adjacent \nii\ line. 
In this
paper we present the results of these observations, comparing the
population of emission line galaxies with the surrounding field and
also our previously published H$\alpha$ survey of the $z=0.31$ cluster 
AC114 (Paper1) \nocite{C+01}.   These observations include 29 galaxies
which were detected in the MIR by \citet{Fadda}, which
provides the opportunity to
compare H$\alpha$ and MIR fluxes for a well defined, volume-limited sample.

The paper is organized as follows.  Our spectroscopic observations with
LDSS++, including the measurement of H$\alpha$ and the statistical correction
for the sampling strategy, are described in \S\ref{sec-ldss}.  The {\it Hubble Space Telescope} ({\it HST}) 
observations
are described in \S\ref{sec-HST}, and the data catalogue is presented in \S\ref{sec-table}.
In \S\ref{sec-results} we present the H$\alpha$ detections and their dependence on cluster
radius, galaxy morphology, and MIR luminosity.  The implications of the
results are discussed in detail in \S\ref{sec-discuss}, and summarized in the conclusions,
\S\ref{sec-conc}.
Throughout the paper we assume a cosmology with $\Omega_\Lambda=0.7, \Omega_m=0.3$, and parameterise the Hubble parameter
as $H_\circ=100h$ km s$^{-1}$ Mpc$^{-1}$.

\section{Observations, Reductions and Analysis}
\subsection{LDSS++ Spectroscopy}\label{sec-ldss}
\subsubsection{Mask design and observations}
The observations were made on the nights of 2001 April 23-24, using
the LDSS$++$ on the AAT.  We use the nod-and-shuffle
technique described in Paper~I\nocite{C+01}, which allows for 
high-precision sky-subtraction in the presence of time-varying 
night-sky emission, and obviates the need for long slits (in order to
get good sampling of the sky adjacent to the object); consequently, it is
possible to work with very small apertures (circular micro-slits of 2\arcsec\ diameter) in the focal
plane mask and thus observe many more objects simultaneously.  The dispersion
is $\sim 2.7$ \AA\ per pixel, with a spectral
resolution $\sim 8$\AA\ FWHM.  As in
Couch et al., we use a blocking filter to restrict our spectral coverage to the
wavelength window in which the \ha\ emission from cluster members
would be seen, $\lambda=7600$\AA--$8000$\AA.  This enables us to overlap microslits
in the wavelength direction, and hence to further increase the multiplex gain.  
The average cluster redshift is $z=0.183$, with a velocity dispersion
of $\sigma\sim1989$ km s$^{-1}$ \citep{SR99}, although this latter value is likely
inflated due to the presence of substructure along the line of sight \citep{Girardi_a1689,CS01}.  The $3-\sigma$ membership
limits are then $z=0.159$--$0.206$.  H$\alpha$ will fall into our band limiting filter
for $z=0.158$--$0.224$; therefore we will detect H$\alpha$ in some background galaxies.

A single LDSS++ mask was created, with galaxy
targets selected from a deep $I$-band image taken with the European Southern
Observatory's 
New Technology Telescope (NTT) in 0\farcs9 seeing conditions.  We first allocated slits
to 34 MIR sources (29 of which turned out to be galaxies), the maximum number possible with a single mask,
comprising 75\% of the sample published by \citet{Fadda}.  
The remaining slits were assigned randomly to galaxies
brighter than $I=22.5$, with higher priority given to brighter galaxies.
In total, we allocate 559 microslits over the full 
$8.7'\times 8.7'$ ($1.1\times 1.1\,h^{-1}\,$Mpc at
$z=0.18$) field-of-view of LDSS$++$.   

The observations were obtained in 
reasonable weather conditions,
with a total exposure time of 19.5 ks, of which half was on the objects and half on
sky.  To overcome the 50\%
losses due to having to obtain sky observations through the mask, we also
experimented with placing mask holes in the position of the galaxy in the
nominal ``sky'' position (5\arcsec\ north of the target).  However, none
of these 76 positions yielded useful additional data, possibly due to guiding
problems when off-target, so we have removed them
from the present analysis.  To ensure that our survey sampled even the most
compact cluster members \citep[e.g.,][]{Drink}, we were deliberately conservative in our star-galaxy
separation, and subsequent comparison with {\it HST} imaging revealed 37 of our
targets to be stars.  Thus, our final sample consists of spectra for 522 galaxies.

The data reduction for nod-and-shuffle observations
is very simple, requiring only that the spectra obtained from
the two nodded positions be subtracted from one another.  
In Paper~I\nocite{C+01} we noted that differences in scattered light across the detector
meant the continuum level in the sky-subtracted object was uncertain, and had
to be corrected for.  For the data in the present paper, we subtracted a surface fit to the background
light on the detector, rejecting light from the slits themselves.  This
removes the scattered light problem, and no further correction was necessary.
Wavelength calibration is based on four lines from a CuAr arc lamp.

\subsubsection{Flux Calibration}\label{sec-flux}
The conversion from detector counts to flux was achieved using the procedure described in
Paper~I\nocite{C+01}.  Nine galaxies with colours typical of early-type cluster members were 
identified, five of which are confirmed  morphologically to be early-types.
The galaxies are relatively isolated, and cover the magnitude range $I=16.4$--$18.4$.
The $I$-band flux within a 2\arcsec\ diameter aperture, corresponding to our
microslit size, is computed from the NTT image, convolved to match the $\sim 2\arcsec$ seeing
of the spectroscopic observations.  This is 
converted to flux units at $\lambda=7700$\AA,
assuming a spectral shape $f_\nu\propto\nu^{-2.2}$, which is consistent with
the observed $V-I\approx1.6$ colours of these galaxies.  This flux is compared with
the average counts per \AA\ in the spectrum, over $\lambda=7700$\AA--$8000$\AA.  
The calibration factor is $(2.0 \pm 0.4) \times 10^{-19}$ ergs~s$^{-1}$~cm$^{-2}$~counts$^{-1}$,
where the uncertainty represents the scatter between the nine galaxies.
Note that this calibration indicates a factor $\sim 4.3$ increase in sensitivity
relative to our AC114 observations (Paper~I), reflecting the superior seeing conditions
and transparency during which the A\,1689 data were obtained.
Measurements are not corrected for aperture effects or dust extinction.
We note that there is no apparent correlation of flux calibration with position on the detector,
over and above the uncertainties of the calibration itself.  

Star formation rates are calculated from H$\alpha$ luminosities $L_{\rm H\alpha}$, using the relation 
$\mbox{SFR} (M_\odot\mbox{yr}^{-1})=2.0\times 10^{-41}L_{\rm H\alpha} (\mbox{ergs~s}^{-1})$
including 1 magnitude of dust extinction \citep{Kenn_review}.  Note that,
throughout this paper, H$\alpha$ fluxes are presented as measured, while
this average, estimated dust correction is applied to the star formation rates.

\subsubsection{\ha\ Detection and measurement}
As in Paper~I\nocite{C+01}, the detection of \ha\ emission lines in our spectra was done manually.  
Each spectrum was examined in its 2-D and optimally extracted 1-D form,
while blinking with the corresponding sky spectrum.  
At the cluster redshift $z=0.181$, H$\alpha$ lies in the middle of a forest
of OH emission lines in the night sky spectrum.  However,
the nod-and-shuffle technique greatly reduces the number
of night sky line residuals that remain over and above expected Poisson noise, after sky subtraction.
In addition, the 2-D images are very useful for distinguishing real emission lines from noise
fluctuations near bright sky lines, as the flux from real features is spread over more pixels.

Each spectrum was first examined and classified independently by MLB and WJC. 
The two lists were then compared, and discrepancies
were re-examined.  The majority of discrepancies were quickly resolved.
Case I galaxies are those with a clear H$\alpha$ detection, and at least
one other emission line to confirm that the identification is correct.  Usually these supporting lines
were one of the \nii$\lambda\lambda$6548,6583 lines; sometimes the [S{\sc ii}]$\lambda\lambda$6717,6731 
lines were also visible.  Emission lines
with no supporting lines, or which could not be confidently distinguished from sky
residuals, are defined as Case II, and are less secure detections.  
Where possible, we compared our detections with the redshifts subsequently
published by \citet{Duc}.  This revealed only one
H$\alpha$ identification (a Case II object) to be incorrect, as it was in fact
an [O{\sc iii}]$\lambda$5007 emission line.  Our remaining detections are consistent with the
\citet{Duc} redshifts where available; in cases where the Duc \etal\ redshift confirms a Case II
detection, we upgrade the quality to Case~I.

We then measure the strength of the \ha\ and adjacent \nii\ lines by fitting a 
line to the continuum blueward and redward of the three lines, and summing the flux above
the continuum over the relevant line bandpass.  Specifically, for H$\alpha$, we compare the flux at
$\lambda=6555$ \AA--$6575$\AA\ with the fit to the continuum at $\lambda=6510$ \AA--$6540$\AA\
and $\lambda=6590$ \AA--$6620$\AA.  For [N{\sc ii}]$\lambda$6583, we define the feature at
$\lambda=6577$ \AA--$6589$\AA, with the same continuum regions as $H\alpha$.
The error vector is computed from the
sky and object spectra, taking into account the gain (0.37 e$^-$ per ADU) and read noise
(1.8 e$^-$ per pixel) of the detector.  We use this error vector to calculate accurate uncertainties
on the measured line fluxes.  No correction is made for stellar absorption.  Typically,
the contribution from absorption is expected to be $\sim <5$ \AA\ \citep{K92,CL_ext}, although the
true correction depends on the past star formation history.  The H$\alpha$ fluxes
are only significantly ($>20$\%) underestimated when the measured equivalent width is comparable
to the absorption correction.

The spectra are sky-noise dominated; thus the {\it r.m.s.} in a feature-free 
region of the continuum gives a reasonable estimate of the noise level,
independent of galaxy flux.  Indeed, we find the {\it r.m.s.} of a random sample of
galaxies to be approximately constant, at $\sim 40$ counts per pixel, although there is some
dependence on wavelength, due to the presence of bright emission lines in the sky
spectrum.  A typical H$\alpha$ emission line
has a FWHM of $\sim 6$ pixels; a $3\sigma$ detection therefore
corresponds to $3\times~40/\sqrt{6}\sim50$ counts above the continuum.  
From the flux calibration computed in \S\ref{sec-flux},
our detection limit is thus approximately 1$\times 10^{-17}$ ergs~s$^{-1}$~cm$^{-2}$,
or 4$h^{-2}\times 10^{38}$ ergs~s$^{-1}$ at $z=0.18$.  From \citet{Kenn_review},
including 1 magnitude of extinction, this flux corresponds to a star formation
rate of 0.008 $h^{-2} M_\odot$~yr$^{-1}$.  We therefore expect that \ha\ emission from galaxies 
with star formation rates comparable 
to that of the Milky Way ($\sim 3 M_\odot$ yr$^{-1}$), will be detected 
unless they are obscured by more than seven magnitudes of extinction at H$\alpha$.
This work represents the deepest H$\alpha$ observations at $z\gg0$ to
date; locally the only observations which are comparable in depth are the very recent
data on Coma and Abell 1367, by \citet{HALF_local}.

\subsubsection{Statistical Weights}\label{sec-field}
To reliably compute global properties of the cluster 
our galaxy sample needs to be weighted by two factors.  The first, which we call the
sampling rate, accounts for the fact that only a fraction
of the galaxies in the $I-$selected sample were observed spectroscopically.  
In addition, some fraction of the galaxies observed, but for which H$\alpha$ was not
detected, will not be cluster members, but rather belong to the foreground or background
field.  This ``membership'' correction is only applicable to galaxies which are not detected
in H$\alpha$, since their redshifts are unknown.  Both of these weights depend on galaxy
$I$-magnitude and projected radius from the cluster centre. 

\begin{figure*}
\centerline{\psfig{file=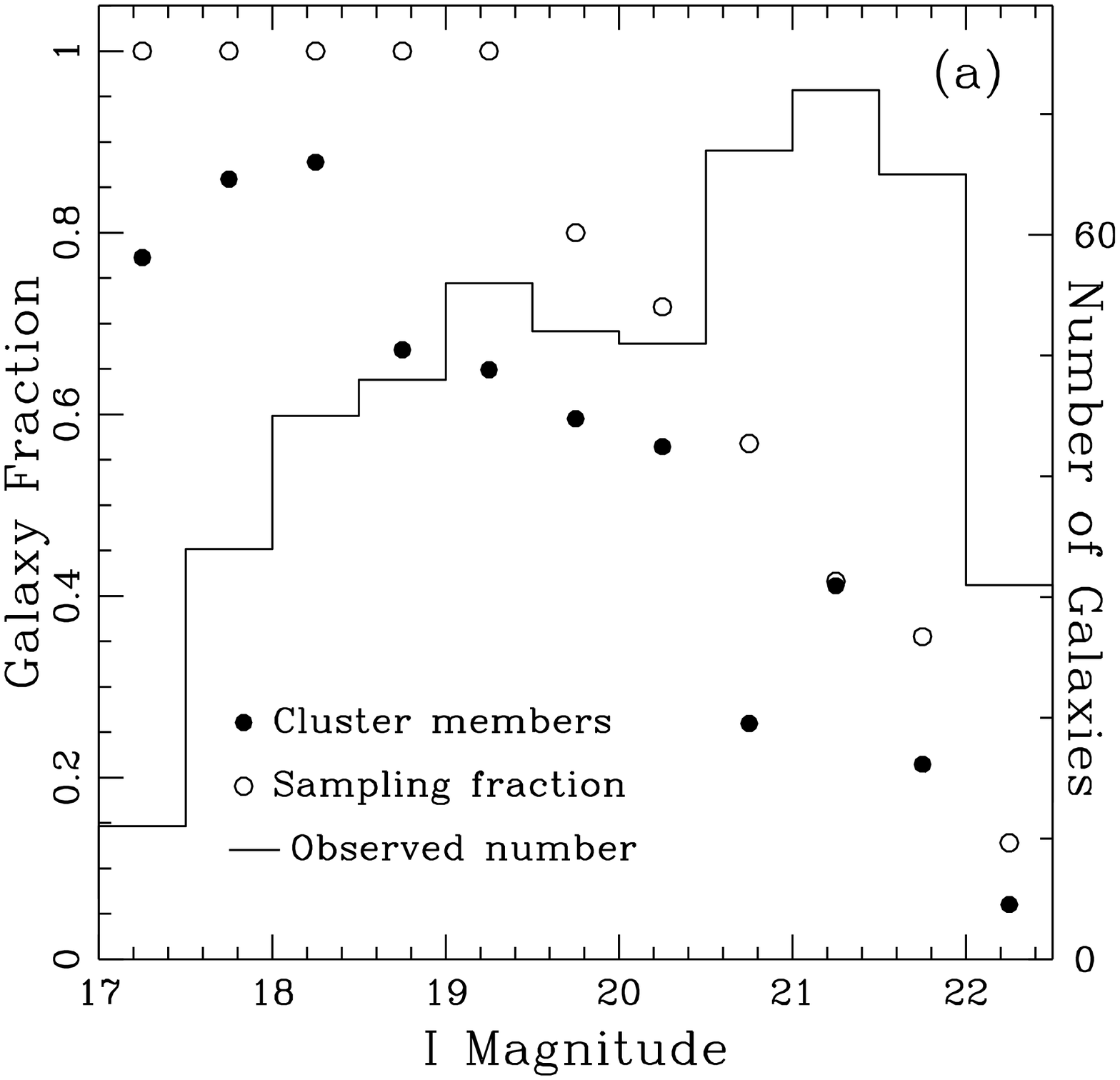,width=2.3in}\hspace*{0.1cm}\psfig{file=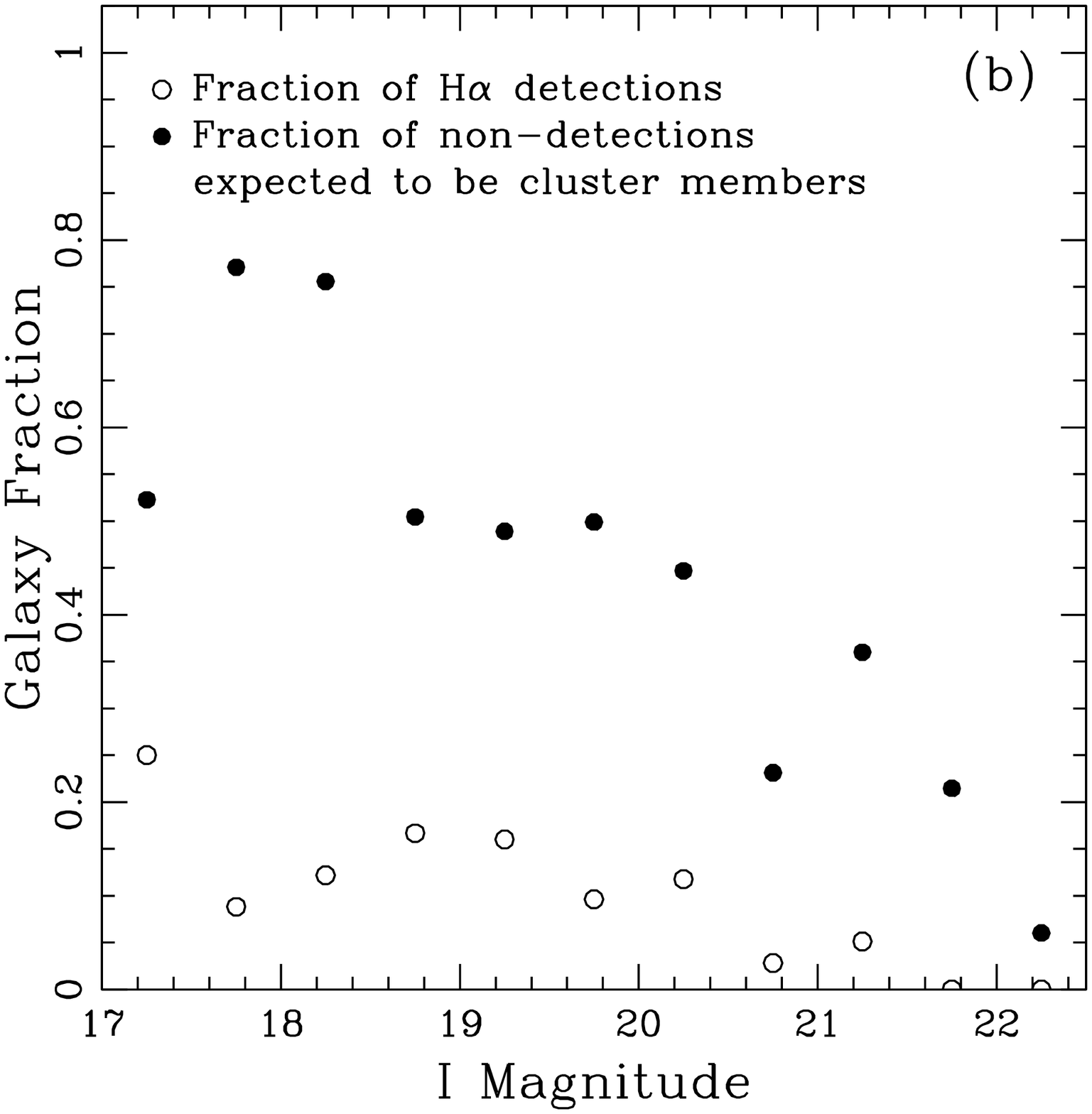,width=2.3in}\hspace*{0.cm}\psfig{file=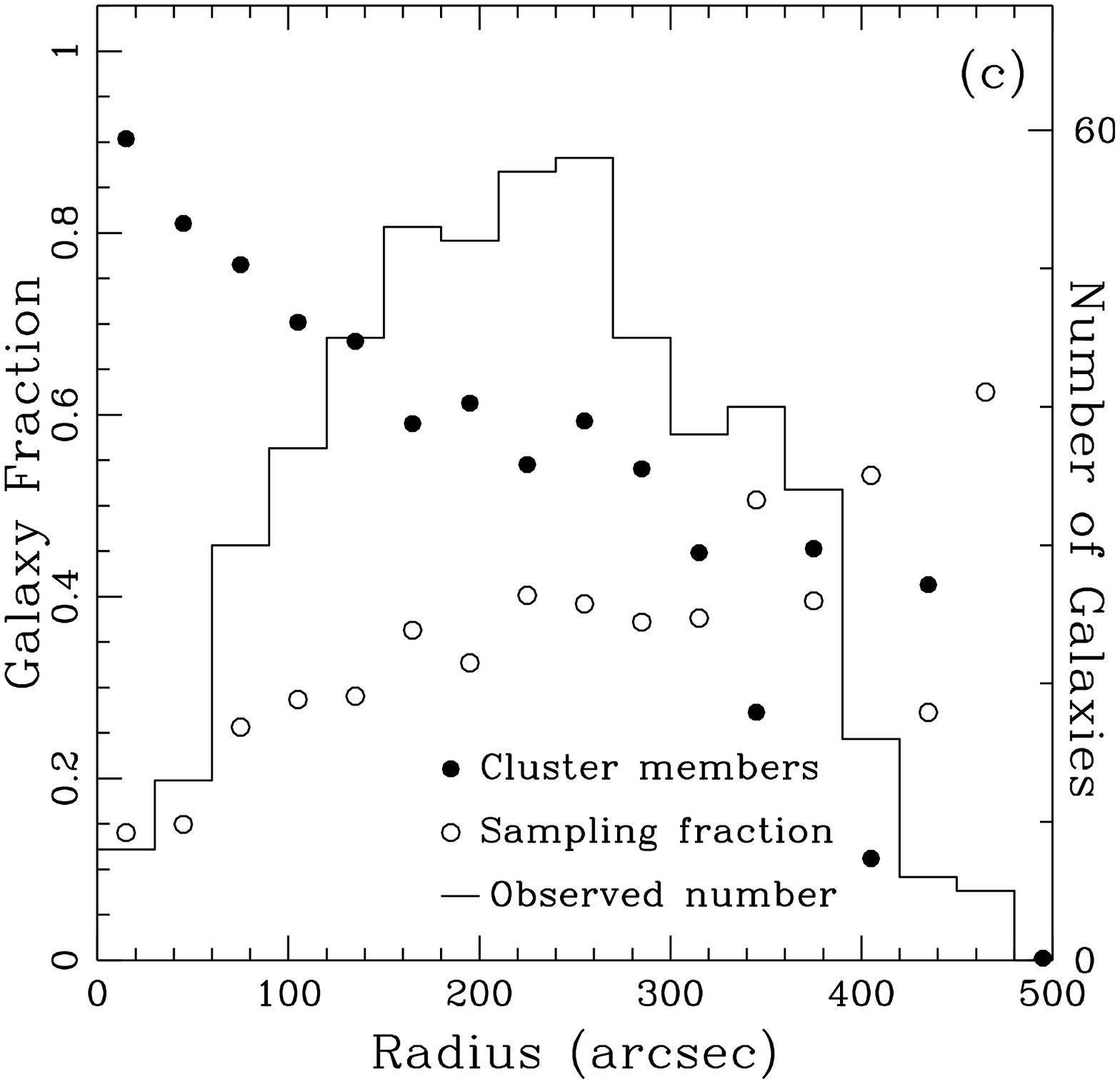,width=2.3in}}
\caption{The determination of luminosity-dependent selection weights applied to the data.
(a)
The {\it open symbols} are the fraction of galaxies in our $I$-band NTT image for
which we obtained a spectrum.  All galaxies
need to be weighted by the inverse of this number to estimate the
corresponding number in the full photometric catalogue.  
{\it Solid symbols} are the fraction of galaxies expected to be
cluster members as a function of $I$ magnitude, computed as the difference between our 
number counts and the field counts of \citet{WHDF5}.   
The total number of
galaxies in the sample is shown as the histogram, with scale on the right
axis.
(b){\it Open symbols} are the fraction of Case I H$\alpha$ detections, as 
a function of $I$-band magnitude.  The {\it solid points} show the difference between
the expected fraction of cluster members and the fraction of H$\alpha$ detections.
Only galaxies which are not detected in H$\alpha$ need to be weighted by this number,
to account for field contamination.
(c)  As panel (a), but for the radial-dependent
selection function.
\label{fig-stat}}
\end{figure*}

The sampling rate as a function of $I$ magnitude is shown in Figure \ref{fig-stat}a.
All galaxies need to be weighted by the inverse of this fraction, to account for the
frequency of such galaxies in a complete, magnitude limited spectroscopic survey.
To determine the membership correction,
we compare the number counts in our $I$-band image with
the field number counts of \citet{WHDF5}.  The expected fraction of cluster members determined
in this way is
also shown in Figure \ref{fig-stat}a.  Some of that expected fraction are actually detected
in H$\alpha$; thus, to determine the fraction
of H$\alpha$-{\it undetected} galaxies which are expected to be cluster members we need to
subtract the fraction of detected galaxies from the expected fraction of cluster members.
We present this in Figure~\ref{fig-stat}b, where the fraction of H$\alpha$-detected galaxies
is shown as a function of $I$-band magnitude.  The solid points show the difference between
the cluster membership fraction and the detected fraction; this gives the fraction of
galaxies undetected in H$\alpha$ that are expected to be members.

The combination of these two luminosity-dependent weights 
(depending on whether or not H$\alpha$ is detected) is called $w_{lum}$.  
A similar weight, $w_{rad}$, is computed to correct for the variation in sampling and membership
with radius, as shown in Figure \ref{fig-stat}c.   We renormalise $w_{rad}$ so that
the sum of the total galaxy weights, $w_{lum}\times w_{rad}$, is equal to the sum of
$w_{lum}$.  

There is potentially another selection bias, as radial annuli are incomplete
at radii beyond $\sim 175$ arcsec, due to the rectangular geometry of the detector.
Thus, galaxies at large radii are underrepresented in the photometric catalogue.
In the absence of a strong radial gradient, however
(see \S\ref{sec-radial}), this does not have a significant effect on our results. 

\subsection{{\it\bf Hubble Space Telescope} Data}\label{sec-HST}
The {\it HST} WFPC2 imaging of A\,1689
was obtained from the ST-ECF archive.  
These data come from the Cycle 5
programme, GO\,5993, and comprise a total of 30 orbits exposure on 15
pointings spread across a $7.5'\times10.0'$ field centred on the
cluster core.    Each of the 15 pointings was observed in two exposures
totalling 2.3\,ks in the F814W ($I_{814}$) passband and a further two
exposures totalling 1.8\,ks through the F606W ($V_{606}$) filter.
These four exposures were each spatially offset by integer pixel
shifts.  Unfortunately, with only two exposures in a given passband it
is difficult to reliably reject cosmic ray events and hot pixels when
processing the images.  For this reason we chose the unusual step of
combining both passbands into a single composite F606W+F814W image.
This combination benefits from the close similarity of galaxies between
the two passbands; for galaxies at the redshift of the cluster the
range in colours is only $V_{606}-I_{814}=0.7$--1.1 for SEDs spanning
Scd to E/S0.  Thus by suitably scaling the two passbands we can combine
them to produce cosmetically clean images for the bulk of the galaxies
used in our analysis.  These data have been previously used in the
analysis of {\it ISOCAM} sources in the field of A\,1689 by \citet{Duc}.

We subsequently calibrate this image onto the $I$-band using our
ground-based imaging, which also provides absolute astrometric
calibration.  The final mosaic covers a field of 69 sq.\ arcmin with
0.17$''$ resolution, a per-pixel exposure time of 4.1\,ks and an
effective point-source sensitivity limit of $I\sim 26$.

The morphologies of the brighter galaxies in the {\it HST} field
($I<21$) have been visually classified by one of us (WJC) onto
the revised Hubble scheme used by the MORPHS project \citep[see ][]{Smail}.
The presence of close neighbours, or signs of interactions such as tidal
tails, were also noted.

\subsection{The Catalogue}\label{sec-table}
Our data and derived quantities are available electronically from {\it MNRAS};
a sample is shown in Table \ref{tab-data}.  
Galaxies are identified by a
unique number (column 1) and their J2000 coordinates (columns 2-3).  The identification numbers
below 50 correspond to the identification numbers in the MIR catalogue of \citet{Fadda}.
The $I-$band magnitude is given in column 4 and,
where H$\alpha$ is detected, we show the quality classification (I or II), redshift and flux 
in columns 5-7.  The corresponding \nii\ flux is given in column 8, and derived star formation 
rates are in column 9.  
Morphological classifications, based on the {\it HST} images, are listed in column 10.
{\scriptsize
\begin{table*} 
\begin{center} 
\caption{\centerline {\sc Sample A\,1689 Data\label{tab-data}}}
\vspace{0.1cm}
\begin{tabular}{lccccccccl} 
\hline\hline
\noalign{\smallskip}
(1)&(2)&(3)&(4)&(5)&(6)&(7)&(8)&(9)&(10)\cr
Id&R.A.&Dec&I&H$\alpha$&Redshift&H$\alpha$ flux&[N{\sc ii}]$\lambda$6583 flux&SFR&Morphology\cr
  &\multispan2{\hfil (J2000)\hfil}&(mag)&Class&&\multispan2{\hfil ($10^{-17}$ ergs s$^{-1}$ cm$^{-2}$)\hfil}&(h$^{-2}$ $M_\odot$ yr$^{-1}$)&\cr
\hline
3& 13 11 25.354 & $-$1 20 37.06 & 17.90& I & 0.192 & $   17.5\pm    6.2$& $   17.0\pm     5.3$& $  0.203\pm   0.072$& Sab\\ 
4& 13 11 27.109 & $-$1 20 58.42 & 18.89& I & 0.214 & $   19.4\pm    6.3$& $    5.1\pm     5.1$& $  0.288\pm   0.094$& Scd\\ 
6& 13 11 27.681 & $-$1 21 07.19 & 19.34& I & 0.215 & $   13.9\pm    6.2$& $   -0.9\pm     4.9$& $  0.208\pm   0.093$& Sd\\ 
7& 13 11 27.821 & $-$1 20 07.76 & 17.69& - & - & -& - & -& E\\ 
...\\
\noalign{\hrule}
\end{tabular}
\end{center} 
\end{table*}
}
\section{Results}\label{sec-results}
\subsection{H$\alpha$-Detected Galaxies}
 \ha\ emission was
detected in 60 of the 522 galaxies observed.  Forty-six of these are Case I detections, 
42 of which lie within $z=0.159$--$0.206$ and are therefore associated with the cluster.
Correcting for sampling frequency and cluster membership (\S\ref{sec-field}), the fraction of Case I
detections is $24\pm4$\%.  This is comparable to the high fraction of blue galaxies brighter than
$R=22.7$ (at least 15\%, possibly as high as 25\%) found by \citet{Duc}.

\begin{figure}
\leavevmode \epsfysize=8cm \epsfbox{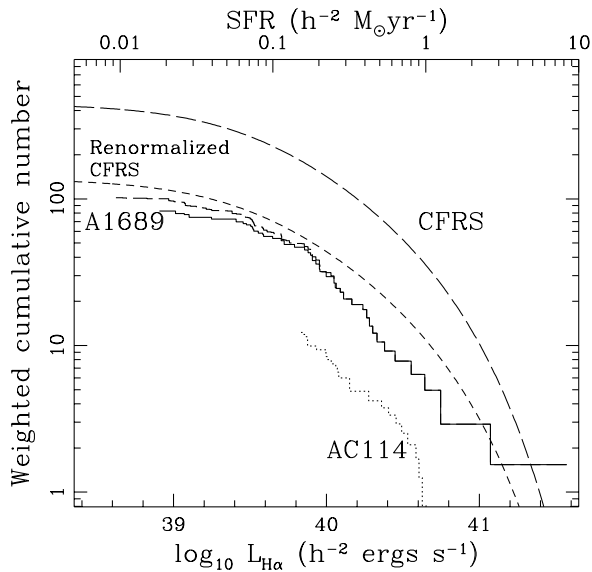}
\caption{The H$\alpha$ luminosity function of A\,1689, compared with that of the CFRS
(TM98).  Both the Case I {\it(solid line)} and II {\it(long dashed line)} functions
are shown.  The CFRS luminosity function is uncorrected for aperture effects and
reddening, to allow a fair comparison with our data.  
We also show this luminosity function renormalized
to account for the different morphological composition of the cluster and field samples,
as described in \S\ref{sec-discuss} ({\it short-dashed line}). 
Finally, we show the luminosity function of AC114, from
Paper~I, renormalized to the number of cluster members in A\,1689 brighter than $I=20.9$,
corresponding to the equivalent luminosity limit in AC114.  
\label{fig-half}}
\end{figure}

The \ha\ luminosity function (H$\alpha$LF) is
shown in Figure \ref{fig-half}, for both Case I and II detections.  
We impose a magnitude limit of $I<22.5$ and adopt a cluster membership
criteria of $z=0.159$--$0.206$.  Applying
the sampling and membership corrections as discussed in \S\ref{sec-field},
to the 504 galaxies observed with $I<22.5$, we expect 328 cluster members
in a complete sample to this depth.  Considering only the Case I detections, the
best fit \citet{S76} parameters of this distribution are $\alpha\sim-0.1$, $L^\ast\sim10^{40.0}$~ergs~s$^{-1}$,
steeper and with a fainter $L^\ast$ than the recently measured local cluster H$\alpha$LF 
\citep[$\alpha\sim -0.7$, $L^\ast\sim10^{40.6}$~ergs~s$^{-1}$, ][]{HALF_local}.

We now compare the cluster H$\alpha$LF with that of the field at $z\sim 0.2$, from the CFRS \citep[][hereafter TM98]{TM98}.
To compute the normalisation, we first note that the luminosity range covered by the
CFRS is similar to ours.  The CFRS samples galaxies with $-23.5<M_{B_{AB}}<-15.5$ mag ($h=1$);
using the typical colour of a Sbc galaxy
at $z=0.2$ \citep{F+95}, this corresponds to $-25.6<M_I<-17.6$.  The faint
end of this limit corresponds to $I=22.5$, which is where we limit our sample.
Over this luminosity range, TM98\nocite{TM98} find
110/131=$84$\% of galaxies have \ha\ flux greater than $3\times 10^{39} h^{-2}$ ergs~s$^{-1}$.  
Thus, we 
normalise the CFRS H$\alpha$LF so that 275 galaxies (84\% of 328) have luminosities greater than 
this limit.  TM98\nocite{TM98} correct their observed fluxes for reddening on a galaxy-by-galaxy
basis, while
we choose to present our data uncorrected for reddening; thus, for comparison we 
reduce the CFRS H$\alpha$ fluxes by a factor of 2.0, corresponding to their average measured
H$\alpha$ extinction.  Finally, TM98\nocite{TM98} also estimate a correction for flux missed
by their 1\farcs75 aperture, which increases their fluxes by a factor which is 1.6 in the mean, 
but which can be as large as a factor of 4.
As we do not correct for aperture effects, we reduce the CFRS H$\alpha$ fluxes by a uniform
factor of 1.6, to make a more fair comparison with our 2\arcsec\ apertures.  
Note that these corrections were not made in Paper~I\nocite{C+01}.
The normalisation of the cluster H$\alpha$LF
is low, relative to that of the field, by a factor of $\sim 5$.  Thus
the total amount of ongoing star formation in this cluster is much less than
in the field, as found in most, if not all, high-density environments \citep[e.g.,][Paper1]{B+97,P+99,Martin00}.

We also show a direct comparison with our previous work on AC114 (Paper~I) in Figure \ref{fig-half}.
Our AC114 observations are statistically complete (after application of a luminosity-dependent
weight to account for membership and selection fractions) to $I=22.25$ ($M_R\sim -16.5+5\log{h}$).  In the 
cosmology we use, this luminosity limit in A\,1689 corresponds to $I=20.9$, so
the latter samples galaxies 1.6 mag less luminous than AC114.  The AC114 H$\alpha$LF is normalized to
a total of 244 cluster members, the weighted number of members in A\,1689 brighter than $I=20.9$.
Only Case I detections are considered. 

The number of galaxies detected in H$\alpha$ is about two times larger in A\,1689, than
in AC114. 
The reason for this difference is not completely understood; both clusters have a large 
fraction of blue galaxies \citep[10-20\%,][]{CS87,Duc}, and any effects of global evolution
would lead to more star formation in the higher redshift cluster AC114, contary to what is observed.
The most likely reason for the difference
is in the dynamical state of the clusters.  AC114 is highly elongated across the sky, but
appears to be fairly relaxed.  On the other hand, although A\,1689 has a smooth, spherical X-ray 
morphology \citep{a1689_xray}, there is clear evidence of substructure in the galaxy
distribution along the line of sight \citep{Girardi_a1689,Duc}, and it appears to be a dynamically
active (i.e. merging) cluster.  
We will explore this difference in detail in later work, when we compare
the H$\alpha$ properties of all four morphologically diverse clusters in our program.

\subsection{The Galaxy Distribution}\label{sec-radial}
\begin{figure}
\leavevmode \epsfysize=8cm \epsfbox{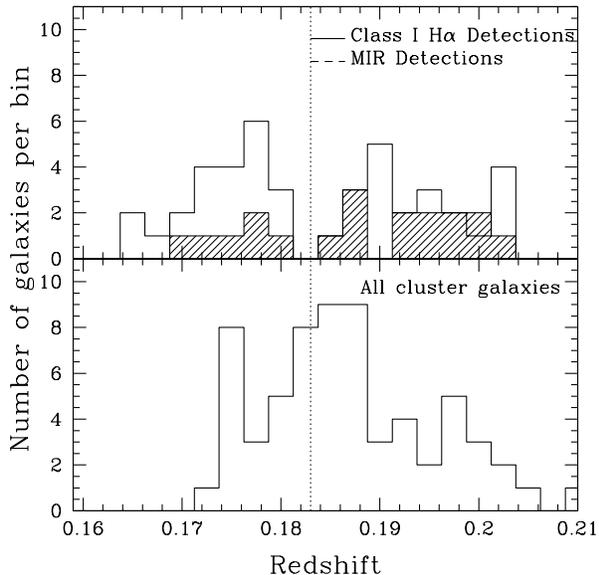}
\caption{{\bf Top: }The {\it open histogram} is the redshift distribution of the Class I H$\alpha$ detections,
while the {\it shaded histogram} is the distribution of the MIR detections from \citet{Fadda}.
{\bf Bottom:} The redshift distribution of the galaxies from the redshift sample
of \citep{Duc}.  The vertical {\it dotted line} is the mean cluster redshift from \citet{SR99}, $z=0.183$.
\label{fig-dynamics}}
\end{figure}

The dynamical properties of the H$\alpha$ population are shown in Figure~\ref{fig-dynamics}, 
where we compare the redshift distribution of the Class I H$\alpha$ detections with that
of the magnitude limited, $V-$selected spectroscopic survey of \citet{Duc}.    First, we point
out that the velocity distribution of the cluster is highly non-Gaussian, with evidence for
substructure in the foreground and background, as previously noted \citep{Girardi_a1689}.  
Thus, the virialized cluster is probably less
massive than naively expected from the very high velocity dispersion
of $\sigma\sim1989$ km s$^{-1}$ \citep{SR99}; from weak lensing measurements, \citet{CS01} find
that the central structure is well fit as an isothermal sphere with $\sigma=1028$ km s$^{-1}$.
Most interestingly, the H$\alpha$ detections do not show a peak at
the mean cluster redshift of $z=0.183$, but rather appear to be associated with the substructure
at $\sim 1600$ km s$^{-1}$ in front of and behind the cluster.
They could either be associated with a galaxy population which is infalling at high
velocity, or with approximately virialized subgroups $\sim20$Mpc in front of and behind the main cluster. 
Either way, these galaxies are not virialized in the main cluster potential,
but are likely to be in the process of being accreted.  This provides good support for models in
which cluster gradients in star formation activity is related to the infall of galaxies
\citep[e.g.,][]{infall,Erica}.   We also show in Figure~\ref{fig-dynamics} the redshift distribution of the galaxies
detected in the MIR \citep{Fadda}.  These galaxies show a similar distribution to the H$\alpha$
sample, suggesting that they are tracing the same population.

\begin{figure}
\leavevmode \epsfysize=8cm \epsfbox{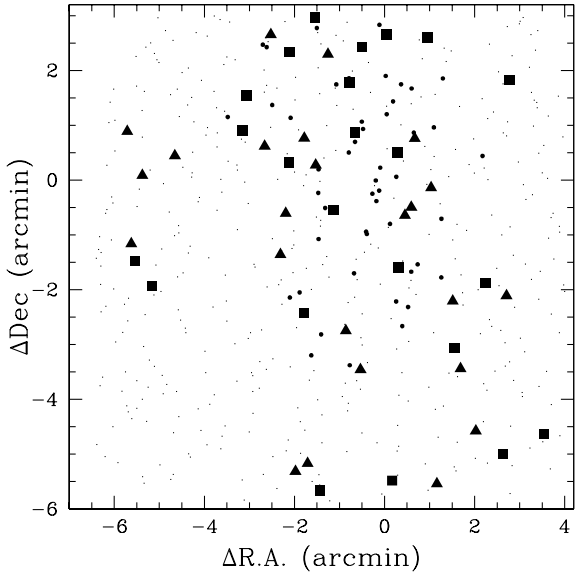}
\caption{The spatial distribution of the targetted galaxies in A\,1689,
brighter than $I=22.5$.  The {\it small dots} indicate the position relative to
the central galaxy for each galaxy for which
a spectrum was obtained.  The {\it filled circles} are those galaxies which are
known to be cluster members, from the spectroscopy of \citet{Duc}.  The {\it large, filled}
symbols are those galaxies we detect at $H\alpha$; the {\it filled squares} are those
with $z<0.185$, while the {\it filled triangles} are those with $z>0.185$.
\label{fig-spatial}}
\end{figure}

The spatial distribution of the targetted galaxies brighter than $I=22.5$ 
is shown in Figure~\ref{fig-spatial}.  The H$\alpha$ detections
in the low- and high-redshift substructures do not appear to correspond to
spatially distinct structures in projected position.  If they do correspond to
foreground/background structures, then, both of these are spread across our full field
of view.

\begin{figure}
\leavevmode \epsfysize=8cm \epsfbox{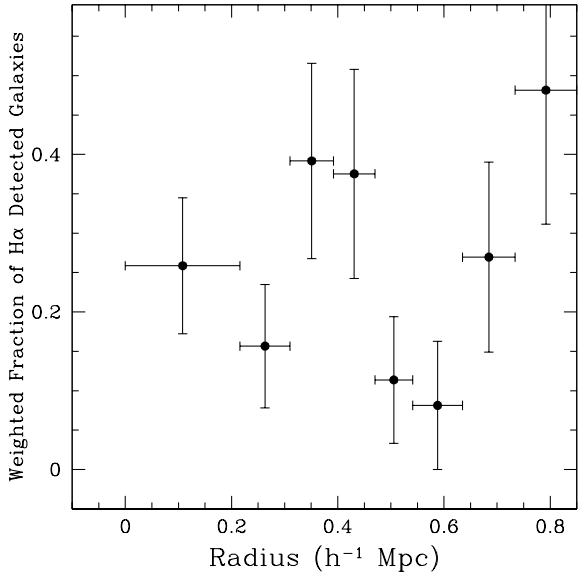}
\caption{The fraction of Case I H$\alpha$ detections as a function of distance from the central
cluster galaxy, for galaxies brighter than $I=22.5$.  The fractions are weighted to account for the varying sampling
frequency and cluster membership as a function of magnitude and radius, as described in the text.  
Radial bins are of varying width, so that each bin includes an equal number ($\sim 60$) of galaxies.
The error bars in the vertical direction
are computed assuming Poisson statistics, and those in the horizontal direction represent the
size of the radial bin.  
\label{fig-radial}}
\end{figure}

In Figure \ref{fig-radial} we show the weighted fraction of Class I H$\alpha$ detections as a function
of distance from the central galaxy (\#20), in Mpc.  
There is no evidence for a trend in the fraction of H$\alpha$ emitters
with radius out to $\sim 1$h$^{-1}$ Mpc.  At all radii the fraction is less than $\sim 50\%$, well below that
of the field at this redshift, $\sim 85$\% (TM98\nocite{TM98}).  A\,1689 is a very massive cluster, with an
estimated virial radius of $r_v\sim 2.8$ Mpc, based on a velocity dispersion of 1028 km s$^{-1}$ \citep{CS01}
and the virial $r_v$-$\sigma$ relation of \citet{Girardi98}.  Thus, these observations are only probing the
core, within 0.4$r_v$, which may explain the lack of a strong gradient in emission line galaxy fraction, 
relative to other studies \citep[e.g.,][]{B+97}.  Furthermore, we reiterate that the emission appears to
be entirely restricted to the structures in the foreground and background of the main cluster, which serves
to dilute any radial gradient present.

\subsection{Galaxy Morphologies}\label{sec-morphs}
{\it HST} imaging is available for 199 of the galaxies in our survey brighter
than $I=21$.
For purposes of presentation we have divided the galaxies into five classes: Elliptical (E),
S0 (including E/S0 and S0/a), Sab (including Sa, Sb, Sab), Scd (including Sc, Sd, Scd and Irr)
and Uncertain (usually because the galaxy lies at the edge of the {\it HST} field).

\begin{figure*}
\centerline{\psfig{file=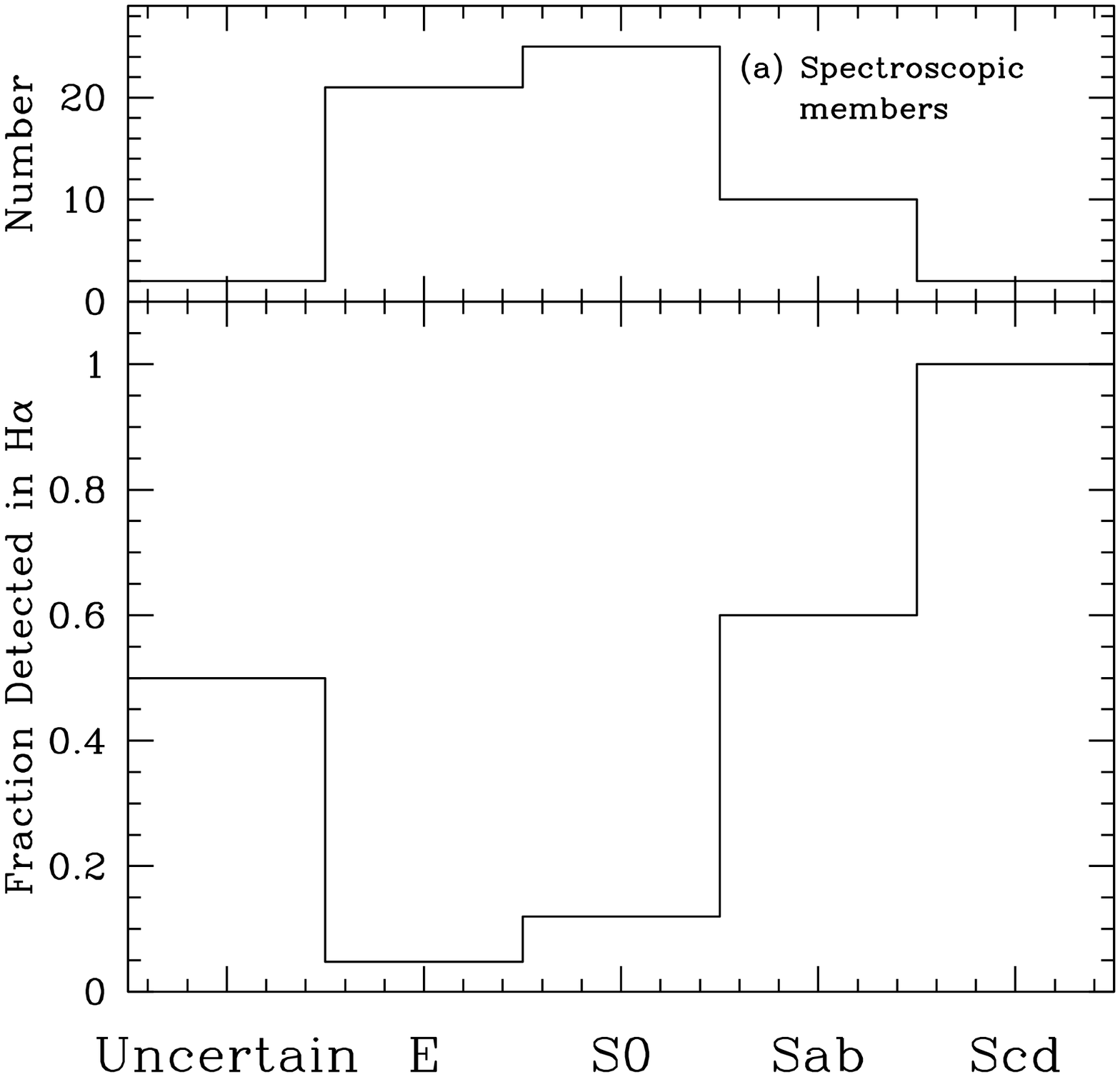,width=2.5in}\hspace*{1cm}\psfig{file=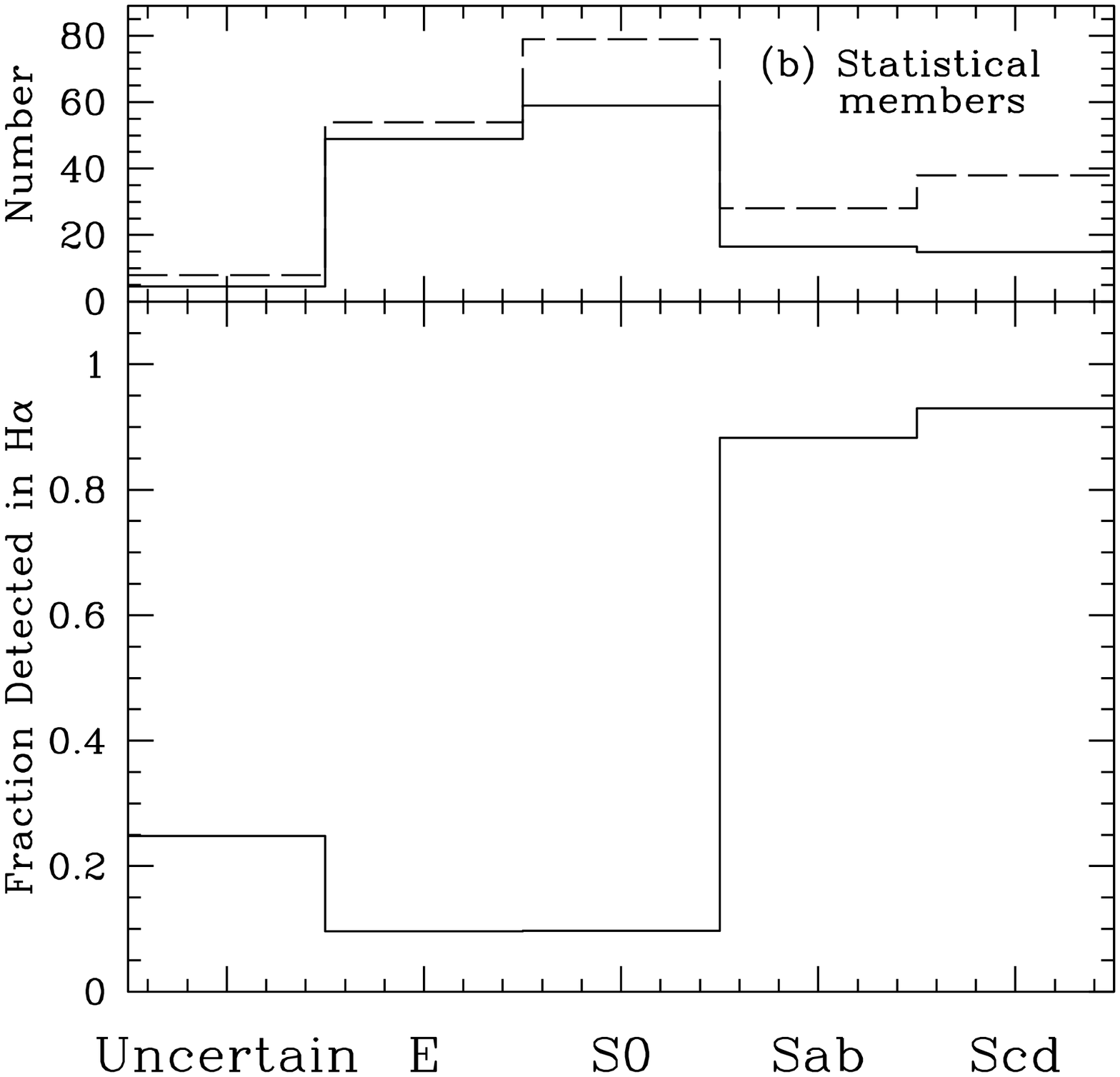,width=2.5in}}
\caption{(a) {\bf Top panel: } The number of galaxies brighter than $I=21$ as a function of Hubble type is shown for confirmed
cluster members, from the spectroscopy of \citet{Duc}.
{\bf Bottom panel:} The
fraction of spectroscopically confirmed cluster galaxies of each type 
securely detected in H$\alpha$.  
(b) Similar to (a) but for the full LDSS++ and {\it HST} sample,
with a statistical field correction as described in the Appendix.
Shown in the top panel is the observed morphological distribution
{\it (dashed line)} and the distribution after making the statistical background
subtraction {\it (solid line)}.  
\label{fig-morph}}
\end{figure*}

To determine the morphological distribution of the cluster population, it is necessary
to correct for contamination from the foreground and background field.  We will consider
two ways of doing this.  The first is to use the redshifts obtained by \citet{Duc}, and
to only consider galaxies within the redshift range $z=0.159$--$0.206$.  This limits our
sample to 60 galaxies.  
The distribution of morphological types among confirmed cluster members is shown in
the top panel of Figure~\ref{fig-morph}a.  The cluster is dominated by early-type galaxies;
only 20$\pm 6$\% of galaxies are of Hubble type Sa or later.  
In the bottom panel we show the fraction of each galaxy type
that is detected in H$\alpha$.  The detections are almost entirely restricted to spiral
galaxies; only 3/46 (6.5\%) of the early-type galaxies are securely detected in H$\alpha$.  

Although the detection 
rate of spiral galaxies appears to be high, the uncertainties are
large because of the small number (12) of late type galaxies in the sample.
In particular, there are only 2 Scd-type
galaxies, so the 100\% detection rate is very uncertain. 
Another approach is to make a statistical correction similar to that
used in  \S\ref{sec-field}, assuming a field morphological composition of 
15\% E, 20\% E/S0, 30\% Sab and 35\% Scd as determined from the Medium Deep Survey
morphological classifications to an equivalent magnitude limit to that considered here \citep{MDS-morph2,GPMC}.  
We describe this procedure more thoroughly in 
the Appendix.  In Figure~\ref{fig-morph}b we show the analogue of Figure~\ref{fig-morph}a,
but now using the larger sample and statistically correcting for the field contribution.
The conclusions are similar, and it is encouraging that the fraction of cluster spiral galaxies 
is 22$\pm 4$\%, in excellent agreement with the fraction determined from the subsample with redshifts.  
The detection rate for \ha\ emission from 
Sab and Scd galaxies is very high, greater than $90$\%.  The fraction of early type galaxies
securely detected in H$\alpha$ remains below 10\%.  We therefore find no difference between
the fraction of late-type cluster galaxies detected in H$\alpha$ and the high fractions
seen in local field samples \citep{K92,Jansen}, a result we discuss in more detail in \S\ref{sec-discuss}.

\subsection{Mid-Infrared Sources}\label{sec-iso}
Of the 522 galaxies targeted, 29 were known MIR sources, from the {\it ISOCAM} observations
of \citet{Fadda}.  Nineteen of
these have measured redshifts $z=0.158$--$0.224$, so any \ha\ emission will lie within our
observable wavelength range (although we note that galaxies at $z>0.206$ are behind 
the cluster as we have defined it).  Another two MIR sources, which did not have known redshifts,
were detected in \ha, bringing the total number of MIR sources which lie in the observable redshift
range to 21.   

\begin{figure}
\leavevmode \epsfysize=8cm \epsfbox{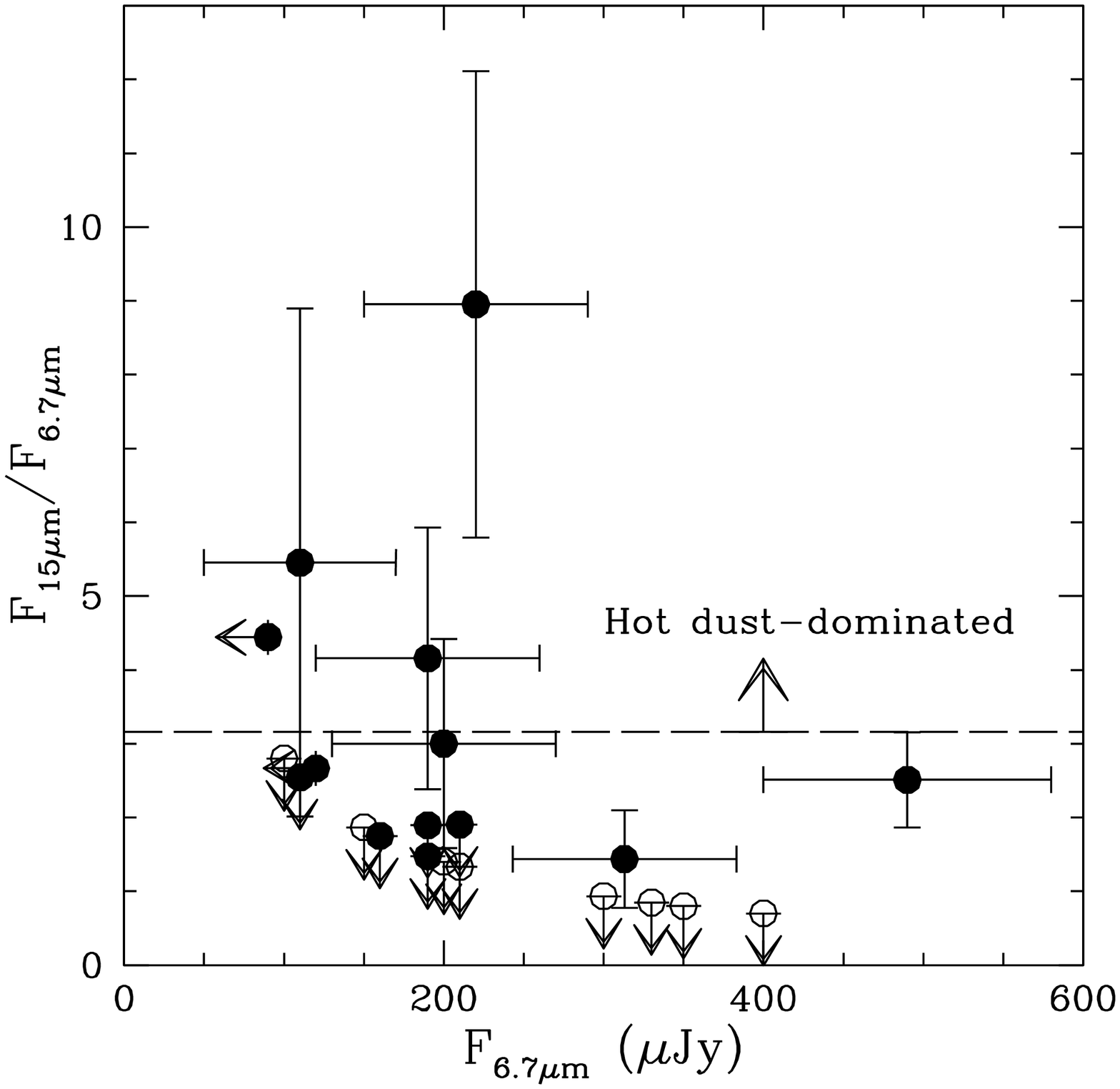}
\caption{The ratio of the MIR flux at \15m\ to the flux at \6m, as a function
of \6m flux for the 21 {\it ISO} sources from \citet{Fadda}, included in the present study.
The {\it filled symbols} represent those galaxies we have detected in H$\alpha$
emission, while the remainder are shown as {\it open circles}.  The horizontal, {\it dashed line} shows an approximate division between
normal spirals (below the line) and galaxies dominated by a hot-dust component \citep{L+00}.
\label{fig-mir}}
\end{figure}

MIR fluxes from \citet{Fadda} are available at  \6m\ and \15m\ wavelengths, 
termed LW2 and LW3.
Emission in the redder band comes predominantly from hot dust while
the \6m\ fluxes are dominated by stellar emission and aromatic carbon compounds related
to star formation in late-type galaxies.  Normal spiral galaxies at $z\sim0.2$ 
are expected to have typical flux
ratios of $L_{15\mu m}/L_{6.7\mu m}<3.2$, while starburst galaxies should have relatively stronger
\15m\ luminosities \citep{Duc,L+00}.
Spectra of dust-enshrouded AGN are also expected to be
dominated by a hot dust component, with aromatic carbon bands similar
to those seen in H{\sc ii} regions \citep{Tran01}.  Although the dust 
temperature and grain size distribution
are expected to be different depending on whether the heating mechanism is star
formation or an AGN, it is not possible to distinguish
the two mechanisms based only on the $L_{15\mu m}/L_{6.7\mu m}$ ratio \citep{L+00}.

\begin{figure*}
\centerline{\psfig{file=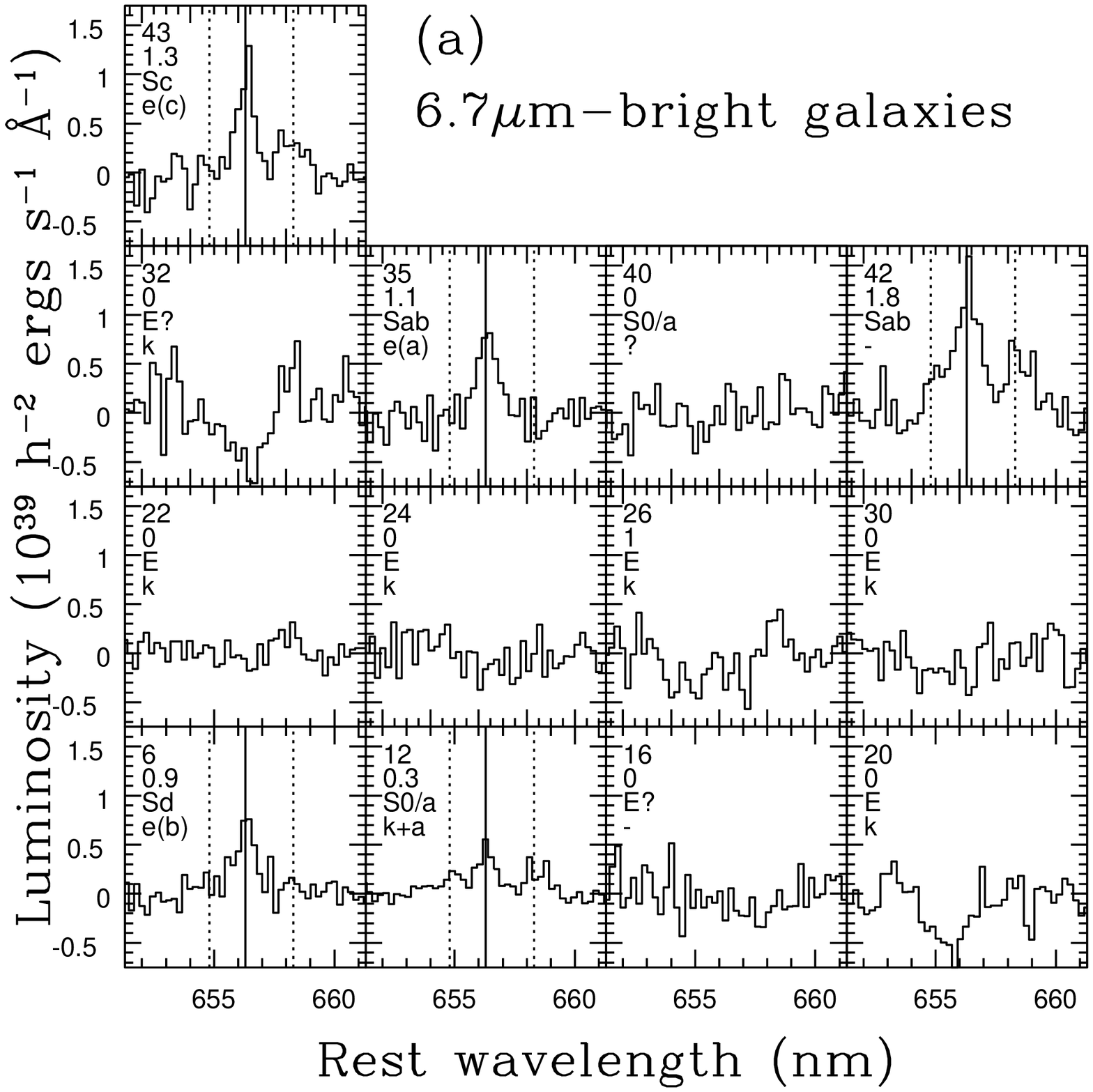,width=3.25in}\hspace*{0.25cm}\psfig{file=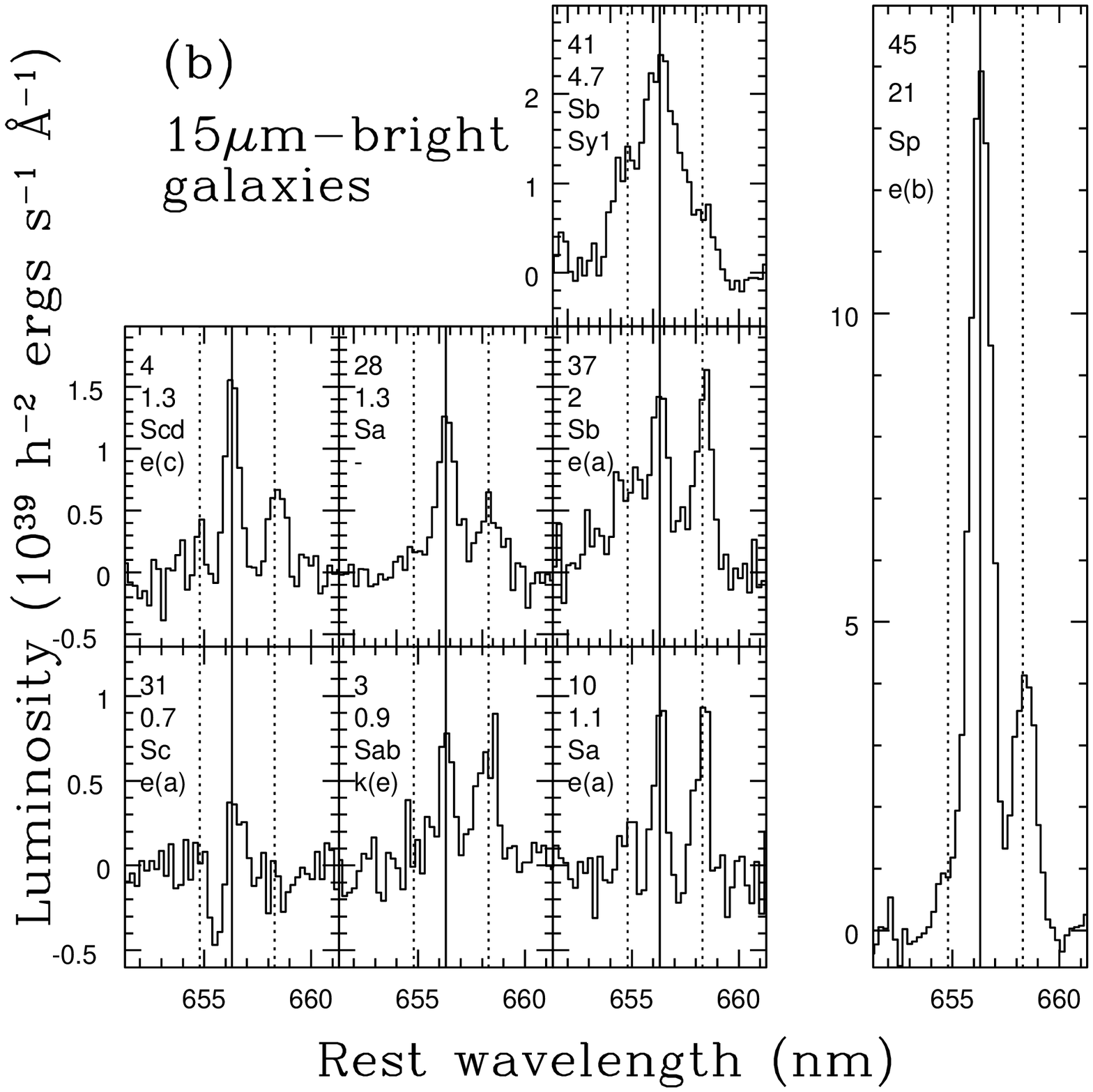,width=3.25in}}
\caption{Continuum-subtracted, rest frame LDSS++ spectra of galaxies detected with{\it ISOCAM}. 
The top row in each box shows the ISO id number of the galaxy.
The next number is the total H$\alpha$ flux in units of $10^{40}$ ergs~s$^{-1}$, followed by
the morphological classification and the 
spectral classification of \citet{Duc}.  A vertical solid line, and
two dashed lines, identify the \ha\ and \nii\ lines, where H$\alpha$ is detected.   
Panel (a) shows the
13 sources detected only at \6m. Panel (b) shows the eight galaxies detected at \15m;
all but two (ISO\# 4, 28) are also detected at \6m.
ISO\#6 and \#4 are associated with a structure behind the cluster at $z=0.215$.
\label{fig-mirspec}}
\end{figure*}

The MIR properties of the 21 {\it ISO} sources observed with LDSS++ are shown in Figure~\ref{fig-mir}.
All but two of these sources are detected at \6m, while only eight are detected at \15m.  
Half of the galaxies detected at \15m\ are unusually bright at \15m, relative to that
expected for normal spiral galaxies.  These spectra are likely dominated
by a hot dust component, although it is not clear from these data alone whether the
dust is heated by star formation or by nuclear activity.
In Figure \ref{fig-mirspec} we show the continuum-subtracted LDSS++ 
spectra of the eight \15m\ sources, and the
13 remaining galaxies detected only at \6m.  Morphological and  spectral classifications
are labelled next to each spectrum.  The spectral classifications are from \citet{Duc}, while
the morphologies are our own classifications.  

\begin{figure}
\leavevmode \epsfysize=8cm \epsfbox{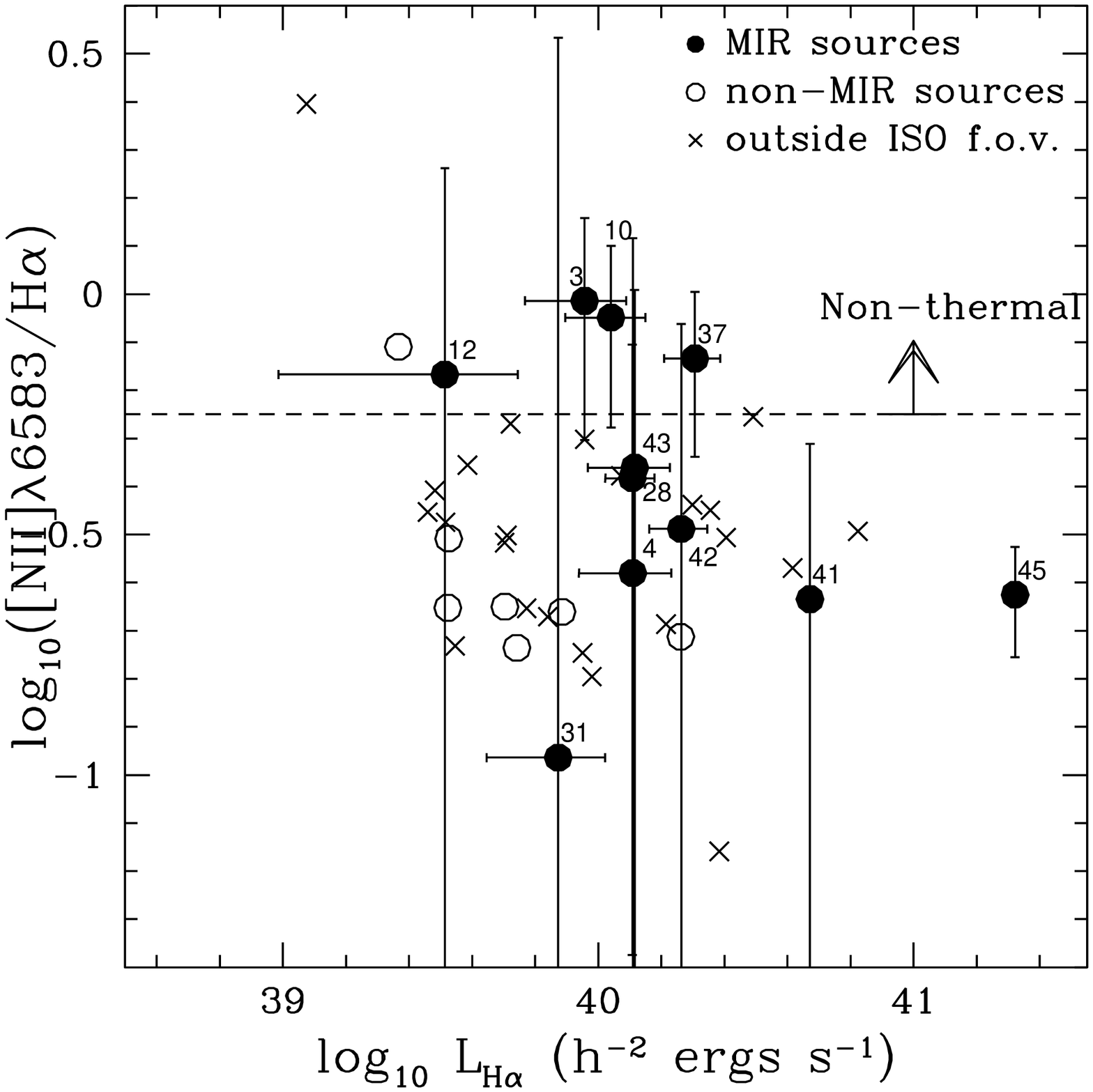}
\caption{The ratio of \nii$\lambda6583$/H$\alpha$, for galaxies with H$\alpha$ and \nii\
flux above our detection limit.  No absorption correction is applied, so these measurements
overestimate the line ratio, particularly for the galaxies \# 31, 37 and 10, which are
known to have strong Balmer absorption.  Galaxies detected with {\it ISOCAM} are shown as
solid points with error bars; the {\it ISO} identification number from \citet{Fadda} is
also shown.  {\it Open symbols} are galaxies which lie within the {\it ISOCAM} field of
view, but are not detected in the MIR.  {\it Crosses} represent galaxies outside the
{\it ISOCAM} field, so no MIR data is available.
The horizontal, {\it dashed line} shows the flux ratio which divides
non-thermal emission sources from line emission due to star formation, as labelled \citep{VO87}.
\label{fig-hanii}}
\end{figure}

Of the 13 MIR galaxies which are only detected at \6m, only 
5 (38.5\%) show H$\alpha$ emission. The remainder are early-type galaxies (E or S0), two
of which clearly show H$\alpha$ in absorption.  Thus it seems likely that the \6m\ light
is dominated by stellar emission, and does not directly trace the hot dust component.
In stark contrast with this, {\it all}
eight of the \15m-detected sources are spiral galaxies securely detected in \ha. 
Thus, our deep H$\alpha$ survey recovers all the potentially-detected star forming galaxies found by
{\it ISOCAM}.

From Figure~\ref{fig-mirspec}b, it is evident
that many of the \15m\ galaxies have atypical spectra in the region of H$\alpha$.
ISO\#41\ has very broad
H$\alpha$ emission and was previously identified as a Seyfert 1 galaxy \citep{Duc}.
Object \#31 has a particularly unusual H$\alpha$ line profile.  H$\alpha$ emission is
detected at 7729\AA, corresponding to a redshift $z=0.1777$; however, the redshift given by
\citet{Duc}, 0.1757, places H$\alpha$ at the absorption trough at 7716\AA.  Therefore, it appears
that the nebular emission is offset from the stellar absorption by $\sim500$~km~s$^{-1}$. 
Three other galaxies, ISO\#3, 10 and 37, show unusully strong \nii$\lambda$6583 emission;
in Figure \ref{fig-hanii} we show the ratio of \nii$\lambda$6583 to H$\alpha$, as a function
of H$\alpha$ luminosity, for all galaxies with H$\alpha$ fluxes above our detection limit.  
Note that the relative
strength of \nii\ and H$\alpha$ are not accurately measured for the broad-lined Seyfert ISO\#41,
as our method is appropriate only for narrow lines. 
Galaxies detected with {\it ISOCAM} are shown with error bars.
\citet{VO87} have shown that galaxies in which \nii$\lambda$6583/$H\alpha>0.55$ 
are almost always associated with non-thermal emission  (i.e., Seyfert, LINERS and narrow
line radio galaxies), rather than with H{\sc ii} regions \citep[see also ][]{K+01}.  Four of the MIR sources (including
the \6m\ source ISO\#12) have ratios
greater than this limit, although the $1\sigma$ error bars overlap this line.  
However, the H$\alpha$ fluxes are not corrected for stellar absorption, and
thus the \nii$\lambda$6583/H$\alpha$ ratio is an overestimate.  In particular, the
galaxy numbers 31, 37 and 10 are known from their full optical spectra to have strong Balmer
absorption lines \citep{Duc}.  These three galaxies are all consistent with \nii$\lambda$6583/H$\alpha<0.55$
if the absorption correction is at least 3\AA, which is not unreasonable.  
Therefore, the only galaxy for which we have unambiguous evidence for a non-thermal contribution
to the H$\alpha$ emission is ISO\#41, which was previously known to be a Seyfert I galaxy \citep{Duc}.

We also show in Figure~\ref{fig-hanii} the H$\alpha$-emitting cluster members
within the {\it ISOCAM} field of view which are undetected in the MIR.
All but one of these has $L_{H\alpha}<10^{40} h^{-2}$~ergs~s$^{-1}$, below the sensitivity 
of the {\it ISOCAM} survey, 0.15 mJy in the \6m\ band.  Thus, the {\it ISOCAM} survey
is nearly complete in the detection of galaxies with H$\alpha$ emission greater than this luminosity
limit, a result we discuss further in \S\ref{sec-discuss} (see Figure~\ref{fig-half2}).

\subsection{Post-starburst Galaxies}
\begin{figure}
\leavevmode \epsfysize=8cm \epsfbox{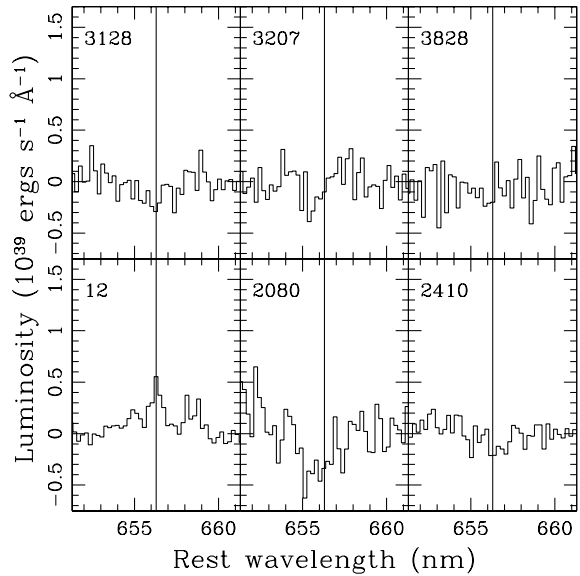}
\caption{Continuum-subtracted, rest-frame LDSS++ spectra of galaxies classified k+a by
\citet{Duc}.  Only galaxy 12 (a \6m\ ISO source) is confidently classified k+a,
and this is the only galaxy detected at H$\alpha$.  The vertical, {\it solid lines}
show the expected position of H$\alpha$ based on the redshifts of \citet{Duc}.
\label{fig-ka}}
\end{figure}

We finally consider the H$\alpha$ properties of those galaxies identified as
k+a by \citet{Duc}.  These are galaxies with little or no \oii\ emission, but
strong Balmer absorption lines which indicate the presence of young stars.  Such
spectra have been interpreted as reflecting a post-starburst state \citep{CS87,P+99,PSG},
or as dust-obscured starbursts \citep{PW00,Smail-radio}.  We observed six galaxies
classified in this way, and we show their continuum-subtracted spectra in the region
of H$\alpha$ in Figure~\ref{fig-ka}.  Only one of these galaxies,
the \6m\ source ISO\#12,  was securely detected in H$\alpha$.  
The H$\alpha$ luminosity of this galaxy is very low, at 
$0.3\times 10^{40} h^{-2}$ ergs s$^{-1}$, corresponding to a 
star formation rate less than $0.1 h^{-2} M_\odot$ yr$^{-1}$.
H$\alpha$ absorption is evident in all the remaining spectra except \# 3828.
In comparison, \citet{A2390_BM} found that at least 2/6 of their k+a galaxies show H$\alpha$
emission.  However, their definition of a k+a galaxy is stricter than that of
\citet{P+99}, which was adopted by \citet{Duc}; furthermore, 
only ISO\#12 is listed as a ``confident'' k+a galaxy by \citet{Duc}, while the remainder
are of  uncertain classification.   Hence, it is difficult to draw strong conclusions;
but it is clear that there is little if any ongoing star formation in these galaxies, although
they must have had at least some star formation in the recent past to give rise to the strong Balmer
absorption.

\section{Discussion}\label{sec-discuss}
Figure \ref{fig-half} demonstrates that, as was the case for the $z=0.31$ cluster
AC114 (Paper~I)\nocite{C+01}, the number of H$\alpha$
emitters in A\,1689 is well below that expected from the surrounding field, by about a factor $\sim 5$.
It is interesting to explore whether or not the difference between  the cluster H$\alpha$LF and
that of the CFRS field can be entirely attributed to the difference in morphological composition.
If we make the reasonable approximation that star formation is restricted to 
galaxies of type Sa or later (see \S\ref{sec-morphs}), then we can
renormalize the CFRS H$\alpha$LF assuming that 65\% of field galaxies at $z\sim0.2$ are of this 
type \citep{MDS-morph2,GPMC}.
In \S\ref{sec-morphs} we determined that $\sim20$\% of the cluster members in our H$\alpha$
survey within the {\it HST} fields are of type Sa or later.  Since the {\it HST} mosaic
covers an area comparable to that of the LDSS++ field, we can appropriately renormalize
the CFRS H$\alpha$LF, as shown as the short-dashed line in Figure \ref{fig-half}.
This renormalization reduces the discrepancy between the field H$\alpha$LF and that of A\,1689,
but does not remove it; H$\alpha$ emitters are still $\gtrsim50$\% more abundant in the field.  This discrepancy
cannot be explained by assuming that star formation, in both the field
and cluster, is mostly restricted to galaxies of type Sc or later.  If this were the case, the
CFRS H$\alpha$LF of Figure~\ref{fig-half} would be $\sim 50$\% higher than assumed in Figure~\ref{fig-half}.

Thus, it appears that the lower star formation
rate of the cluster galaxies is not due only to the difference in morphological composition,
relative to the field.  The next step is to look for evidence that H$\alpha$ emission
in galaxies of a given morphological type is lower in clusters, relative to the field
\citep[e.g.,][]{B+98}.  We will do this by comparing our data with the local field galaxy
sample of  \citet{Jansen}.  Our sample is 50\% complete to $I=21$ (see Figure~\ref{fig-stat}a), which
corresponds approximately to $M_B=-15.9+5\log{h}$, using the colour transformations in
\citet{F+95}.  There are 173 galaxies in the sample of \citet{Jansen} brighter than
this limit, and we take this as our comparison sample.  

If we adopt the statistical background subtraction method
and consider the results of the full sample (Figure~\ref{fig-morph}b), the
fraction of late-type galaxies in A\,1689 detected in H$\alpha$ is $\gtrsim 90$\% although
this fraction is not very well determined, due to the small number of late-type galaxies and
the uncertainty of the background correction. 
In the sample of \citet{Jansen}, 104 of the 121 galaxies brighter than $M_B=-15.9+5\log{h}$
classified as Sa or later show H$\alpha$ emission with $W_\circ(H\alpha)>2$\AA.  
This fraction, 86\%, is in excellent agreement with our results for A\,1689.  Thus, we do not 
find evidence that the few spiral galaxies
in A\,1689 are less likely to have H$\alpha$ emission, relative to their field counterparts.
 
Turning our attention to the early-type galaxy population, Jansen et al. find
that 5/17 of their ellipticals and 7/25 of their S0 galaxies brighter than $M_B=-15.9+5\log{h}$ show
$W_\circ(H\alpha)>2$\AA.  Thus, 28\% of their early-type galaxies show H$\alpha$ emission,
compared with less than 10\% in A\,1689 cluster early-types (Figure~\ref{fig-morph}).  
Even rejecting the three early-type galaxies in Jansen et al.'s sample with Seyfert nuclei, the
fraction of early-type galaxies with H$\alpha$ emission is still twice as large as
the fraction we find in A\,1689.

These results suggest that the difference between the A\,1689 H$\alpha$LF and that of the CFRS,
corrected to the same morphological composition, may be due to the fact that cluster E and S0 galaxies exhibit
H$\alpha$ emission less frequently than their field counterparts.  This cannot be tested
directly, as we do not have morphological classifications of the CFRS sample, and we depend
upon the MDS sample of \citet{MDS-morph2} being complete and representative, relative to the CFRS sample
of TM98.  However, we note that our results are in  agreement with
independent evidence from absorption line analysis that suggests that field early-type galaxies have younger
stellar populations on average than their cluster counterparts \citep{A+98,Trager,Kunt02}.  Our results could also 
be consistent with those of \citet{B+98}, who
showed that disk-dominated galaxies in clusters have lower [O{\sc ii}] emission than analogous
galaxies in the field, if a substantial number of those disk-dominated galaxies are S0. 
On the other hand, \citet{P+99} found that the low star formation rates in clusters at $z\sim 0.4$ 
could be attributed to a large population of passive
spiral galaxies.  Our results do not appear to be consistent with such a population, although this
is strongly dependent on the statistical subtraction of the field.  Using only the spectroscopic
sample of \citet{Duc} we find that only 60\% of Sab galaxies are detected in H$\alpha$; it would
be interesting to see if this holds for a larger sample of confirmed members.

\begin{figure}
\leavevmode \epsfysize=8cm \epsfbox{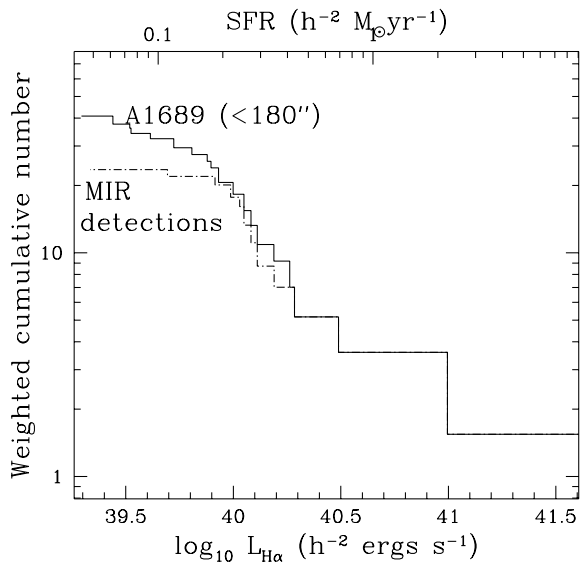}
\caption{The H$\alpha$ luminosity function of A\,1689, restricted to the area covered by
the ISOCAM survey of \citet{Fadda}.  We compare this with the H$\alpha$ luminosity function
of those galaxies detected in the MIR. 
\label{fig-half2}}
\end{figure}

We now consider the claims of \citet{Fadda} and \citet{Duc}, that much of the star formation
in A\,1689 is obscured by dust, and detectable in the MIR.  
We first reiterate that our observations are very sensitive
to low \ha\ fluxes, and we would detect a galaxy with a star formation rate
greater than $\sim 3 M_\odot$ yr$^{-1}$ unless it was obscured by more than 7 magnitudes of dust. 
We show the H$\alpha$LF of our full sample, restricted to the 3$\arcmin$ field of the ISOCAM
survey of \citet{Fadda}, in Figure~\ref{fig-half2}.  We compare this with the H$\alpha$LF 
of galaxies detected in the MIR, within the same area.  This demonstrates that the MIR catalogue is approximately
complete to H$\alpha$ luminosities $L_{H\alpha}>10^{40} h^{-2}$~ergs~s$^{-1}$.  However, our
H$\alpha$ survey is at least four times more sensitive and detects twice as many 
star forming galaxies in this volume (which contribute an additional $16$\% 
to the total amount of star formation).  
  
\begin{figure}
\leavevmode \epsfysize=8cm \epsfbox{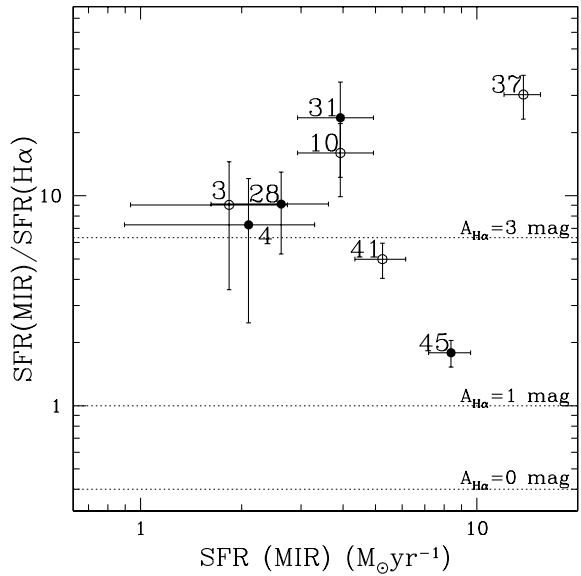}
\caption{A comparison of star formation rates estimated from the H$\alpha$ emission
line and those determined from \15m\ fluxes \citep{Duc}.  The H$\alpha$-derived star
formation rates assume an extinction of 1 magnitude.  {\it Open symbols} are
those galaxies with either broad H$\alpha$ emission, or \nii/H$\alpha>0.55$, possibly indicative
of significant non-thermal contribution, in which case both star formation rates are 
overestimates.  The ISO identification number is shown next to each point.
The horizontal, {\it dotted lines} show the expected offset if the extinction is
0, 1 and 3 magnitudes at 6563\AA.
\label{fig-sfr}}
\end{figure}

In Figure \ref{fig-sfr} we compare star formation rates measured from H$\alpha$ with
those measured from \15m\ luminosities, as estimated by \citet{Duc}.  We have increased the published
MIR star formation rates by a factor of 1.3 to account for the different cosmologies
assumed.  For those MIR sources with a flux ratio $F_{15}/F_{K^\prime}\leq0.2$ we
reduce the MIR-derived star formation rate by a factor of two, to correct for photospheric 
emission\footnote{Duc et al. actually recommend this correction be made for galaxies with $F_{15}/F_{K^\prime}\leq0.1$,
but the only two other sources affected by our more generous limits are actually upper limits
on the flux ratio.}.

If the dust giving rise to the MIR emission is heated predominantly by star formation,
then our direct comparison with the H$\alpha$ fluxes supports
the conclusions of \citet{Duc} and \citet{Fadda}, that optical line
emission underestimates the star formation in \15m\ sources.  Star formation rates estimated
from the H$\alpha$ line (which include a correction for 1 magnitude of dust extinction) are
lower than those derived from the \15m\ data by a factor of between 2 and 20 for most of 
the galaxies.  If we were to correct the observed H$\alpha$ fluxes for an assumed $5$\AA\ of
stellar absorption, this would reduce the discrepancy to a factor of 1.7--10.
The effect is more severe for star formation rates determined from the \oii\
emission line, where the underestimate is a factor of 10 to 100 \citep{Duc}.

This suggests that some of the MIR-detected galaxies could be obscured by up to 3 magnitudes of
extinction at 6563\AA.
The main remaining difficulty is to understand how much 
of the MIR emission in the {\it ISOCAM} sample
arises from non-thermal activity.  For galaxies of this luminosity, a 30\% contribution
to their bolometric infrared luminosity 
from AGN-heating is not unreasonable \citep{Tran01}; in this case, the SFR estimated
from both the MIR luminosity and the H$\alpha$ luminosity are overestimated.  
However, even if galaxies are bolometrically dominated by
star formation, the bulk of their \15m\ emission could still be coming from an
AGN, especially in a MIR-selected sample \citep[but see][]{S+95,CE01}. 
There are currently no published wide-field, local
MIR surveys which can directly address this question, although such studies are underway
\citep[e.g., ][]{AA01,ELAIS}.  Preliminary results from the ELAIS
survey suggests that only $\sim 30$\% of \15m\ sources are AGN-dominated, while 45\% are starbursts and
15\% are absorption line galaxies \citep{ELAIS}.  However, this mix is likely to be strongly redshift-dependent,
and the AGN contribution at $z\sim0.2$ depends on which of the AGN or starburst population is
evolving more strongly with redshift \citep{RE99}. 
For at least one of the eight \15m-detected galaxies in our H$\alpha$ sample (ISO\# 41) there is a clear contribution
to the dust heating from an AGN source, and this could hold for up to half of the sample
(including ISO\# 3,10, 37), depending on the uncertain correction for stellar absorption and the 
usefulness of the \nii$\lambda6583$/H$\alpha$ ratio as an indicator of non-thermal emission.

\section{Conclusions}\label{sec-conc}
We have obtained LDSS++ spectroscopy for 522 galaxies over a field covering $\sim 1.1\times1.1 h^{-1}$ Mpc
in the z=0.18 cluster A\,1689.  We find the following results:
\begin{itemize}
\item 24$\pm4$\% of cluster members more luminous than $M_R=-16.5+5\log{h}$, 
statistically weighted to account for sampling-related 
selection effects, are confidently detected in H$\alpha$ emission above a flux limit of
4$h^{-2}\times 10^{38}$ ergs~s$^{-1}$, corresponding to a limiting
star formation rate of 0.008 $h^{-2}$ $M_\odot$ yr$^{-1}$,
assuming 1 magnitude of extinction.

\item The normalization of the cluster H$\alpha$ luminosity function ($z=0.18$)
is a factor of $\sim 5$ lower than that of
the field at $z\approx0.2$, from TM98\nocite{TM98}, and a factor of $\sim 2$ higher than that in our previous
study of AC114 ($z=0.31$).

\item From the 199 galaxies for which we have {\it HST} {\it WFPC2} morphologies, as well as LDSS++
spectroscopy, we determine that $\sim$20\% of the cluster members (regardless of their H$\alpha$ properties) 
are of type Sa or later, much less 
than the $\sim 65\%$ expected in a sample of field galaxies.
Like their field counterparts, these spiral galaxies have a high incidence of H$\alpha$
detection, $\gtrsim 90$\% .  On the other hand,
H$\alpha$ emission in early-type galaxies (E and S0) is less frequently observed than in similar
galaxies in the field, based on the sample of \citet{Jansen}.

\item After renormalizing the CFRS H$\alpha$LF (TM98\nocite{TM98}) to
account for the difference in morphological composition between the field and cluster samples
a $\sim 50$\% deficit of cluster galaxies with H$\alpha$ emission remains.  Combined with the high
incidence of emission in cluster late-type galaxies, this suggests that the star formation
rates of early-type cluster galaxies are suppressed, relative to their field counterparts.

\item All of the galaxies detected in the \15m\ mid-infrared band with {\it ISOCAM} are detected in
H$\alpha$.  The MIR survey of \citet{Fadda} selects an approximately complete
sample of H$\alpha$-emitting galaxies with $L_{H\alpha}>10^{40} h^{-2}$~ergs~s$^{-1}$.  Our
deep H$\alpha$ survey probes $\sim4$ times fainter than this, doubling the sample of 
detected star-forming galaxies and recovering an
additional $\sim16$\% of all H$\alpha$ emission in the cluster.

\item The population of H$\alpha$-emitting galaxies in A\,1689 appears to be 
associated with substructure in the immediate foreground and background of the cluster.  The MIR
detections trace the dynamics of the H$\alpha$ population, suggesting that they are related.

\item Star formation rates derived from \15m\ emission, neglecting a significant
non-thermal component, are a factor 2--20 larger than those estimated
from H$\alpha$, assuming 1 magnitude of extinction.  This difference
is reduced to a factor $\lesssim 10$ if we account for an assumed 5\AA\ underlying
stellar absorption.  Both star formation rate estimates may be
overestimated if there is a significant contribution to the dust heating from
AGN, for which there may be evidence in up to 50\% of the eight \15m\ sources we observe
in H$\alpha$.

\end{itemize}

In summary, the star forming galaxies in the core of A\,1689 are significantly rarer than
in the surrounding field.  This may be due primarily to a suppression of star formation
in S0 or elliptical galaxies.
Our deep H$\alpha$ survey finds all \15m\ MIR sources catalogued in this region, so there is no evidence for any star
formation completely hidden at H$\alpha$, although the estimated star formation rates from H$\alpha$
may be too low by assuming standard extinctions.  This depends, however, on the largely
unknown contribution of AGN-heated dust to the MIR flux.

%
%

\section*{Acknowledgements}
We would like to thank P.-A. Duc, D. Fadda and B. Poggianti for 
providing their data prior to publication,
and for making many useful suggestions regarding the comparison with the {\it ISOCAM}
data.  We also thank Jean-Paul Kneib for providing the NTT $I-$band image
from which the sample was selected.  Finally, we thank an anonymous
referee for helpful criticisms.
MLB is supported by a PPARC rolling grant for extragalactic astronomy
and cosmology at the University of Durham.  
WJC acknowledges funding support from the Australian Research Council.
IRS acknowledges support from a Royal Society URF and a Phillip Leverhulme
Prize Fellowship.
Based on observations made with the NASA/ESA Hubble Space Telescope,
obtained from the data archive at the Space Telescope Institute. STScI
is operated by the association of Universities for Research in
Astronomy, Inc. under the NASA contract  NAS 5-26555.

\bibliographystyle{apj}
\bibliography{ms}

\begin{thebibliography}{70}
\expandafter\ifx\csname natexlab\endcsname\relax\def\natexlab#1{#1}\fi

\bibitem[{{Abraham} {et~al.}(1999){Abraham}, {Ellis}, {Fabian}, {Tanvir}, \&
  {Glazebrook}}]{A+98}
{Abraham}, R.~G., {Ellis}, R.~S., {Fabian}, A.~C., {Tanvir}, N.~R., \&
  {Glazebrook}, K. 1999, MNRAS, 303, 641

\bibitem[{{Abraham} {et~al.}(1996){Abraham}, {van den Bergh}, {Glazebrook},
  {Ellis}, {Santiago}, {Surma}, \& {Griffiths}}]{MDS-morph2}
{Abraham}, R.~G., {van den Bergh}, S., {Glazebrook}, K., {Ellis}, R.~S.,
  {Santiago}, B.~X., {Surma}, P., \& {Griffiths}, R.~E. 1996, ApJS, 107, 1

\bibitem[{{Alexander} {et~al.}(2001){Alexander}, {Brandt}, {Hornschemeier},
  {Garmire}, {Schneider}, {Bauer}, \& {Griffiths}}]{A+01}
{Alexander}, D.~M., {Brandt}, W.~N., {Hornschemeier}, A.~E., {Garmire}, G.~P.,
  {Schneider}, D.~P., {Bauer}, F.~E., \& {Griffiths}, R.~E. 2001, AJ, 122, 2156

\bibitem[{{Almaini} {et~al.}(1999){Almaini}, {Lawrence}, \& {Boyle}}]{ALB}
{Almaini}, O., {Lawrence}, A., \& {Boyle}, B.~J. 1999, MNRAS, 305, L59

\bibitem[{{Aussel} \& {Alexander}(2001)}]{AA01}
{Aussel}, H. \& {Alexander}, D.~M. 2001, American Astronomical Society Meeting,
  198

\bibitem[{{Balogh} \& {Morris}(2000)}]{A2390_BM}
{Balogh}, M.~L. \& {Morris}, S.~L. 2000, MNRAS, 318, 703

\bibitem[{Balogh {et~al.}(1997)Balogh, Morris, Yee, Carlberg, \&
  Ellingson}]{B+97}
Balogh, M.~L., Morris, S.~L., Yee, H. K.~C., Carlberg, R.~G., \& Ellingson, E.
  1997, ApJL, 488, 75

\bibitem[{Balogh {et~al.}(1999)Balogh, Morris, Yee, Carlberg, \&
  Ellingson}]{PSG}
---. 1999, ApJ, 527, 54

\bibitem[{{Balogh} {et~al.}(2000){Balogh}, {Navarro}, \& {Morris}}]{infall}
{Balogh}, M.~L., {Navarro}, J.~F., \& {Morris}, S.~L. 2000, ApJ, 540, 113

\bibitem[{Balogh {et~al.}(1998)Balogh, Schade, Morris, Yee, Carlberg, \&
  Ellingson}]{B+98}
Balogh, M.~L., Schade, D., Morris, S.~L., Yee, H. K.~C., Carlberg, R.~G., \&
  Ellingson, E. 1998, ApJL, 504, 75

\bibitem[{{Blain} {et~al.}(1999){Blain}, {Smail}, {Ivison}, \&
  {Kneib}}]{Blain99}
{Blain}, A.~W., {Smail}, I., {Ivison}, R.~J., \& {Kneib}, J.-P. 1999, MNRAS,
  302, 632

\bibitem[{{Bower}(1991)}]{PSext}
{Bower}, R.~G. 1991, MNRAS, 248, 332

\bibitem[{Butcher \& Oemler(1984)}]{BO84}
Butcher, H. \& Oemler, A. 1984, ApJ, 285, 426

\bibitem[{{Charlot} {et~al.}(2002){Charlot}, {Kauffmann}, {Longhetti},
  {Tresse}, {White}, {Maddox}, \& {Fall}}]{CL_ext}
{Charlot}, S., {Kauffmann}, G., {Longhetti}, M., {Tresse}, L., {White},
  S.~D.~M., {Maddox}, S.~J., \& {Fall}, S.~M. 2002, MNRAS, 330, 876

\bibitem[{{Chary} \& {Elbaz}(2001)}]{CE01}
{Chary}, R. \& {Elbaz}, D. 2001, ApJ, 556, 562

\bibitem[{{Clowe} \& {Schneider}(2001)}]{CS01}
{Clowe}, D. \& {Schneider}, P. 2001, A\&A, 379, 384

\bibitem[{{Couch} {et~al.}(2001){Couch}, {Balogh}, {Bower}, {Smail},
  {Glazebrook}, \& {Taylor}}]{C+01}
{Couch}, W.~J., {Balogh}, M.~L., {Bower}, R.~G., {Smail}, I., {Glazebrook}, K.,
  \& {Taylor}, M. 2001, ApJ, 549, 820

\bibitem[{Couch {et~al.}(1998)Couch, Barger, Smail, Ellis, \& Sharples}]{C+98}
Couch, W.~J., Barger, A.~J., Smail, I., Ellis, R.~S., \& Sharples, R.~M. 1998,
  ApJ, 497, 188

\bibitem[{Couch \& Sharples(1987)}]{CS87}
Couch, W.~J. \& Sharples, R.~M. 1987, MNRAS, 229, 423

\bibitem[{{Cowie} {et~al.}(1999){Cowie}, {Songaila}, \& {Barger}}]{Cowie+99}
{Cowie}, L.~L., {Songaila}, A., \& {Barger}, A.~J. 1999, AJ, 118, 603

\bibitem[{Dressler {et~al.}(1997)Dressler, Oemler, Couch, Smail, Ellis, Barger,
  Butcher, Poggianti, \& Sharples}]{D+97}
Dressler, A., Oemler, A., Couch, W.~J., Smail, I., Ellis, R.~S., Barger, A.,
  Butcher, H.~R., Poggianti, B.~M., \& Sharples, R.~M. 1997, ApJ, 490, 577

\bibitem[{{Drinkwater} {et~al.}(1999){Drinkwater}, {Phillipps}, {Gregg},
  {Parker}, {Smith}, {Davies}, {Jones}, \& {Sadler}}]{Drink}
{Drinkwater}, M.~J., {Phillipps}, S., {Gregg}, M.~D., {Parker}, Q.~A., {Smith},
  R.~M., {Davies}, J.~I., {Jones}, J.~B., \& {Sadler}, E.~M. 1999, ApJL, 511,
  L97

\bibitem[{{Duc} {et~al.}(2002){Duc}, {Poggianti}, {Fadda}, {Elbaz}, {Flores},
  {Chanial}, {Franceschini}, {Moorwood}, \& {Cesarsky}}]{Duc}
{Duc}, P.-A., {Poggianti}, B.~M., {Fadda}, D., {Elbaz}, D., {Flores}, H.,
  {Chanial}, P., {Franceschini}, A., {Moorwood}, A., \& {Cesarsky}, C. 2002,
  A\&A, 382, 60

\bibitem[{{Durret} {et~al.}(1994){Durret}, {Gerbal}, {Lachieze-Rey},
  {Lima-Neto}, \& {Sadat}}]{a1689_xray}
{Durret}, F., {Gerbal}, D., {Lachieze-Rey}, M., {Lima-Neto}, G., \& {Sadat}, R.
  1994, A\&A, 287, 733

\bibitem[{{Dwarakanath} \& {Owen}(1999)}]{DO99}
{Dwarakanath}, K.~S. \& {Owen}, F.~N. 1999, AJ, 118, 625

\bibitem[{{Ellingson} {et~al.}(2001){Ellingson}, {Lin}, {Yee}, \&
  {Carlberg}}]{Erica}
{Ellingson}, E., {Lin}, H., {Yee}, H.~K.~C., \& {Carlberg}, R.~G. 2001, ApJ,
  547, 609

\bibitem[{{Fabian} \& {Iwasawa}(1999)}]{FI}
{Fabian}, A.~C. \& {Iwasawa}, K. 1999, MNRAS, 303, L34

\bibitem[{{Fadda} {et~al.}(2000){Fadda}, {Elbaz}, {Duc}, {Flores},
  {Franceschini}, {Cesarsky}, \& {Moorwood}}]{Fadda}
{Fadda}, D., {Elbaz}, D., {Duc}, P.-A., {Flores}, H., {Franceschini}, A.,
  {Cesarsky}, C.~J., \& {Moorwood}, A.~F.~M. 2000, A\&A, 361, 827

\bibitem[{{Fadda} {et~al.}(2002){Fadda}, {Flores}, {Hasinger}, {Franceschini},
  {Altieri}, {Cesarsky}, {Elbaz}, \& {Ferrando}}]{Fadda02}
{Fadda}, D., {Flores}, H., {Hasinger}, G., {Franceschini}, A., {Altieri}, B.,
  {Cesarsky}, C.~J., {Elbaz}, D., \& {Ferrando}, P. 2002, A\&A, 383, 838

\bibitem[{{Flores} {et~al.}(1999){Flores}, {Hammer}, {Thuan}, {C{\' e}sarsky},
  {Desert}, {Omont}, {Lilly}, {Eales}, {Crampton}, \& {Le F{\`
  e}vre}}]{Flores99}
{Flores}, H., {Hammer}, F., {Thuan}, T.~X., {C{\' e}sarsky}, C., {Desert},
  F.~X., {Omont}, A., {Lilly}, S.~J., {Eales}, S., {Crampton}, D., \& {Le F{\`
  e}vre}, O. 1999, ApJ, 517, 148

\bibitem[{Fukugita {et~al.}(1995)Fukugita, Shimasaku, \& Ichikawa}]{F+95}
Fukugita, M., Shimasaku, K., \& Ichikawa, T. 1995, PASP, 107, 945

\bibitem[{{Genzel} {et~al.}(1998){Genzel}, {Lutz}, {Sturm}, {Egami}, {Kunze},
  {Moorwood}, {Rigopoulou}, {Spoon}, {Sternberg}, {Tacconi-Garman}, {Tacconi},
  \& {Thatte}}]{Genzel98}
{Genzel}, R., {Lutz}, D., {Sturm}, E., {Egami}, E., {Kunze}, D., {Moorwood},
  A.~F.~M., {Rigopoulou}, D., {Spoon}, H.~W.~W., {Sternberg}, A.,
  {Tacconi-Garman}, L.~E., {Tacconi}, L., \& {Thatte}, N. 1998, ApJ, 498, 579

\bibitem[{{Girardi} {et~al.}(1997){Girardi}, {Fadda}, {Escalera}, {Giuricin},
  {Mardirossian}, \& {Mezzetti}}]{Girardi_a1689}
{Girardi}, M., {Fadda}, D., {Escalera}, E., {Giuricin}, G., {Mardirossian}, F.,
  \& {Mezzetti}, M. 1997, ApJ, 490, 56

\bibitem[{{Girardi} {et~al.}(1998){Girardi}, {Giuricin}, {Mardirossian},
  {Mezzetti}, \& {Boschin}}]{Girardi98}
{Girardi}, M., {Giuricin}, G., {Mardirossian}, F., {Mezzetti}, M., \&
  {Boschin}, W. 1998, ApJ, 505, 74

\bibitem[{{Glazebrook} \& {Bland-Hawthorn}(2001)}]{GBH}
{Glazebrook}, K. \& {Bland-Hawthorn}, J. 2001, PASP, 113, 197

\bibitem[{{Glazebrook} {et~al.}(1995){Glazebrook}, {Peacock}, {Miller}, \&
  {Collins}}]{GPMC}
{Glazebrook}, K., {Peacock}, J.~A., {Miller}, L., \& {Collins}, C.~A. 1995,
  MNRAS, 275, 169

\bibitem[{{Iglesias-P{\' a}ramo} {et~al.}(2002){Iglesias-P{\' a}ramo},
  {Boselli}, {Cortese}, {V{\' i}lchez}, \& {Gavazzi}}]{HALF_local}
{Iglesias-P{\' a}ramo}, J., {Boselli}, A., {Cortese}, L., {V{\' i}lchez},
  J.~M., \& {Gavazzi}, G. 2002, A\&A, 384, 383

\bibitem[{{Jansen} {et~al.}(2000){Jansen}, {Fabricant}, {Franx}, \&
  {Caldwell}}]{Jansen}
{Jansen}, R.~A., {Fabricant}, D., {Franx}, M., \& {Caldwell}, N. 2000, ApJS,
  126, 331

\bibitem[{{Kennicutt}(1998)}]{Kenn_review}
{Kennicutt}, R.~C., J. 1998, ARA\&A, 36, 189

\bibitem[{Kennicutt(1992)}]{K92}
Kennicutt, R.~C. 1992, ApJ, 388, 310

\bibitem[{{Kewley} {et~al.}(2001){Kewley}, {Dopita}, {Sutherland}, {Heisler},
  \& {Trevena}}]{K+01}
{Kewley}, L.~J., {Dopita}, M.~A., {Sutherland}, R.~S., {Heisler}, C.~A., \&
  {Trevena}, J. 2001, ApJ, 556, 121

\bibitem[{{Kodama} \& {Bower}(2001)}]{KB01}
{Kodama}, T. \& {Bower}, R.~G. 2001, MNRAS, 321, 18

\bibitem[{{Kuntschner} {et~al.}(2002){Kuntschner}, {Smith}, {Colless},
  {Davies}, {Kaldare}, \& {Vazdekis}}]{Kunt02}
{Kuntschner}, H., {Smith}, R.~J., {Colless}, M., {Davies}, R.~L., {Kaldare},
  R., \& {Vazdekis}, A. 2002, MNRAS, submitted

\bibitem[{{La Franca} {et~al.}(2002){La Franca}, {Matute}, {Fiore},
  {Gruppioni}, {Pozzi}, {Vignali}, \& {HELLAS, ELAIS collaborations}}]{ELAIS}
{La Franca}, F., {Matute}, I., {Fiore}, F., {Gruppioni}, C., {Pozzi}, F.,
  {Vignali}, C., \& {HELLAS, ELAIS collaborations}. 2002, in ASP Conference
  Series, ed. R.~Maiolino, A.~Marconi, \& N.~Nagar, Vol. astro-ph/0109308

\bibitem[{{Laurent} {et~al.}(2000){Laurent}, {Mirabel}, {Charmandaris},
  {Gallais}, {Madden}, {Sauvage}, {Vigroux}, \& {Cesarsky}}]{L+00}
{Laurent}, O., {Mirabel}, I.~F., {Charmandaris}, V., {Gallais}, P., {Madden},
  S.~C., {Sauvage}, M., {Vigroux}, L., \& {Cesarsky}, C. 2000, A\&A, 359, 887

\bibitem[{Lilly {et~al.}(1995)Lilly, Tresse, Hammer, Crampton, \&
  LeF\`{e}vre}]{CFRS6}
Lilly, S.~J., Tresse, L., Hammer, F., Crampton, D., \& LeF\`{e}vre, O. 1995,
  ApJ, 455, 108

\bibitem[{{Lutz} {et~al.}(1998){Lutz}, {Spoon}, {Rigopoulou}, {Moorwood}, \&
  {Genzel}}]{Lutz98}
{Lutz}, D., {Spoon}, H.~W.~W., {Rigopoulou}, D., {Moorwood}, A.~F.~M., \&
  {Genzel}, R. 1998, ApJL, 505, L103

\bibitem[{Madau {et~al.}(1996)Madau, Ferguson, Dickinson, Giavalisco, Steidel,
  \& Fruchter}]{Madau}
Madau, P., Ferguson, H.~C., Dickinson, M.~E., Giavalisco, M., Steidel, C., \&
  Fruchter, A. 1996, MNRAS, 283, 1388

\bibitem[{{Margoniner} {et~al.}(2001){Margoniner}, {de Carvalho}, {Gal}, \&
  {Djorgovski}}]{Margo}
{Margoniner}, V.~E., {de Carvalho}, R.~R., {Gal}, R.~R., \& {Djorgovski}, S.~G.
  2001, ApJL, 548, L143

\bibitem[{{Martin} {et~al.}(2000){Martin}, {Lotz}, \& {Ferguson}}]{Martin00}
{Martin}, C.~L., {Lotz}, J., \& {Ferguson}, H.~C. 2000, ApJ, 543, 97

\bibitem[{{Marzke} {et~al.}(1994){Marzke}, {Geller}, {Huchra}, \&
  {Corwin}}]{marzke_cfa_morph}
{Marzke}, R.~O., {Geller}, M.~J., {Huchra}, J.~P., \& {Corwin}, H.~G. 1994, AJ,
  108, 437

\bibitem[{{Metcalfe} {et~al.}(2001){Metcalfe}, {Shanks}, {Campos}, {McCracken},
  \& {Fong}}]{WHDF5}
{Metcalfe}, N., {Shanks}, T., {Campos}, A., {McCracken}, H.~J., \& {Fong}, R.
  2001, MNRAS, 323, 795

\bibitem[{{Metcalfe} {et~al.}(1991){Metcalfe}, {Shanks}, {Fong}, \&
  {Jones}}]{MSFJ}
{Metcalfe}, N., {Shanks}, T., {Fong}, R., \& {Jones}, L.~R. 1991, MNRAS, 249,
  498

\bibitem[{Poggianti {et~al.}(1999)Poggianti, Smail, Dressler, Couch, Barger,
  Butcher, Ellis, \& Oemler}]{P+99}
Poggianti, B.~M., Smail, I., Dressler, A., Couch, W.~J., Barger, A.~J.,
  Butcher, H., Ellis, R.~S., \& Oemler, A. 1999, ApJ, 518, 576

\bibitem[{{Poggianti} \& {Wu}(2000)}]{PW00}
{Poggianti}, B.~M. \& {Wu}, H. 2000, ApJ, 529, 157

\bibitem[{{Rigopoulou} {et~al.}(2000){Rigopoulou}, {Franceschini}, {Aussel},
  {Genzel}, {van der Werf}, {Cesarsky}, {Dennefeld}, {Oliver},
  {Rowan-Robinson}, {Mann}, {Perez-Fournon}, \& {Rocca-Volmerange}}]{ISO-HDF}
{Rigopoulou}, D., {Franceschini}, A., {Aussel}, H., {Genzel}, R., {van der
  Werf}, P., {Cesarsky}, C.~J., {Dennefeld}, M., {Oliver}, S.,
  {Rowan-Robinson}, M., {Mann}, R.~G., {Perez-Fournon}, I., \&
  {Rocca-Volmerange}, B. 2000, ApJL, 537, L85

\bibitem[{{Roche} \& {Eales}(1999)}]{RE99}
{Roche}, N. \& {Eales}, S.~A. 1999, MNRAS, 307, 111

\bibitem[{{Roussel} {et~al.}(2001){Roussel}, {Sauvage}, {Vigroux}, \&
  {Bosma}}]{RSVB}
{Roussel}, H., {Sauvage}, M., {Vigroux}, L., \& {Bosma}, A. 2001, A\&A, 372,
  427

\bibitem[{{Rowan-Robinson} {et~al.}(1997){Rowan-Robinson}, {Mann}, {Oliver},
  {Efstathiou}, {Eaton}, {Goldschmidt}, {Mobasher}, {Serjeant}, {Sumner},
  {Danese}, {Elbaz}, {Franceschini}, {Egami}, {Kontizas}, {Lawrence},
  {MCMahon}, {Norgaard-Nielsen}, {Perez-Fournon}, \&
  {Gonzalez-Serrano}}]{RR+97}
{Rowan-Robinson}, M., {Mann}, R.~G., {Oliver}, S.~J., {Efstathiou}, A.,
  {Eaton}, N., {Goldschmidt}, P., {Mobasher}, B., {Serjeant}, S. B.~G.,
  {Sumner}, T.~J., {Danese}, L., {Elbaz}, D., {Franceschini}, A., {Egami}, E.,
  {Kontizas}, M., {Lawrence}, A., {MCMahon}, R., {Norgaard-Nielsen}, H.~U.,
  {Perez-Fournon}, I., \& {Gonzalez-Serrano}, J.~I. 1997, MNRAS, 289, 490

\bibitem[{Schechter(1976)}]{S76}
Schechter, P. 1976, ApJ, 203, 297

\bibitem[{Smail {et~al.}(1997)Smail, Dressler, Couch, Ellis, Oemler, Butcher,
  \& Sharples}]{Smail}
Smail, I., Dressler, A., Couch, W.~J., Ellis, R.~S., Oemler, A., Butcher, H.,
  \& Sharples, R.~M. 1997, ApJS, 110, 213

\bibitem[{{Smail} {et~al.}(1999){Smail}, {Morrison}, {Gray}, {Owen}, {Ivison},
  {Kneib}, \& {Ellis}}]{Smail-radio}
{Smail}, I., {Morrison}, G., {Gray}, M.~E., {Owen}, F.~N., {Ivison}, R.~J.,
  {Kneib}, J.-P., \& {Ellis}, R.~S. 1999, ApJ, 525, 609

\bibitem[{{Spinoglio} {et~al.}(1995){Spinoglio}, {Malkan}, {Rush}, {Carrasco},
  \& {Recillas-Cruz}}]{S+95}
{Spinoglio}, L., {Malkan}, M.~A., {Rush}, B., {Carrasco}, L., \&
  {Recillas-Cruz}, E. 1995, ApJ, 453, 616

\bibitem[{{Struble} \& {Rood}(1999)}]{SR99}
{Struble}, M.~F. \& {Rood}, H.~J. 1999, ApJS, 125, 35

\bibitem[{{Trager} {et~al.}(2000){Trager}, {Faber}, {Worthey}, \& {Gonz{\'
  a}lez}}]{Trager}
{Trager}, S.~C., {Faber}, S.~M., {Worthey}, G., \& {Gonz{\' a}lez}, J.~J.~.
  2000, AJ, 120, 165

\bibitem[{{Tran} {et~al.}(2001){Tran}, {Lutz}, {Genzel}, {Rigopoulou}, {Spoon},
  {Sturm}, {Gerin}, {Hines}, {Moorwood}, {Sanders}, {Scoville}, {Taniguchi}, \&
  {Ward}}]{Tran01}
{Tran}, Q.~D., {Lutz}, D., {Genzel}, R., {Rigopoulou}, D., {Spoon}, H.~W.~W.,
  {Sturm}, E., {Gerin}, M., {Hines}, D.~C., {Moorwood}, A.~F.~M., {Sanders},
  D.~B., {Scoville}, N., {Taniguchi}, Y., \& {Ward}, M. 2001, ApJ, 552, 527

\bibitem[{{Tresse} \& {Maddox}(1998)}]{TM98}
{Tresse}, L. \& {Maddox}, S.~J. 1998, ApJ, 495, 691

\bibitem[{{Veilleux} \& {Osterbrock}(1987)}]{VO87}
{Veilleux}, S. \& {Osterbrock}, D.~E. 1987, ApJS, 63, 295

\bibitem[{Whitmore {et~al.}(1993)Whitmore, Gilmore, \& Jones}]{WGJ}
Whitmore, B.~C., Gilmore, D.~M., \& Jones, C. 1993, ApJ, 407, 489

\bibitem[{{Wilman} {et~al.}(2000){Wilman}, {Fabian}, \& {Gandhi}}]{WFG}
{Wilman}, R.~J., {Fabian}, A.~C., \& {Gandhi}, P. 2000, MNRAS, 318, L11

\end{thebibliography}

\appendix
\section{A statistical, morphologically-dependent background correction}\label{sec-app}
The redshifts of most galaxies in our sample which are not detected in H$\alpha$ are unknown; therefore a statistical
correction needs to be applied to account for the fact that some of these will be field
galaxies in the foreground and background of the cluster.  In \S\ref{sec-field} we computed
this correction for the global population by comparing the number counts in our field with
those of \citet{MSFJ}.  However, when we consider galaxies of a fixed morphological type,
we need to consider the different morphological
composition of the field.  We take this from the {\it Medium Deep Survey} 
\citep[MDS, ][]{MDS-morph2,GPMC}.  The MDS comprises visual morphological classifications
of a magnitude-limited field sample, from a survey of serendipitous {\it WFPC2} fields.
These classifications are on the same scheme to that used here, as shown by 
comparisons in \citet{Smail}.  As in \S\ref{sec-morphs}, we will consider four
classes: Elliptical (E), S0 (including E/S0 and S0/a), Sab (including Sa, Sb, Sab)
and Scd (including Sc, Sd, Scd and Irr).
Over the magnitude range $I=17$--$21$, the field
population is composed of 15\% E, 20\% S0, 30\% Sab and 35\% Scd, with only a small
dependence on limiting magnitude.  This is in reasonably good agreement with the distributions 
for the local field from \citet{WGJ} (18\% E, 23\% S0 and 59\% later types) and \citet{marzke_cfa_morph}
(10\% E, 30\% S0, 40\% Sab, 20\% Scd at M$^\ast$).

Using the total field contamination
estimate from Figure~\ref{fig-stat}a, we can then determine the fraction of galaxies of each
morphological type which are expected to be cluster members. 
From this number
we subtract the fraction of galaxies of that type which we detect in H$\alpha$, yielding
the fraction of undetected galaxies of that type which are expected cluster members.  
We show some sample numbers relevant to this
calculation in Table~\ref{tab-fieldcor}.  The first three columns show the galaxy type,
the number and fraction of that type in the sample (ignoring the Uncertain
class), and the number of galaxies detected in H$\alpha$.  The rest of the calculation
is best illustrated by a couple of examples.   For this purpose, we will assume the
field contributes 40\% to the sample, which is the case at 
$I\sim19$ (Figure~\ref{fig-stat}a).  Using the population mix presented above,
we calculate that 6\% of all galaxies in our sample are field ellipticals; this number is shown in
column 4.  We measure that 27\%
of our galaxies are ellipticals (column 2); therefore, we expect $(27-6)/27=78\%$ of these to be cluster members,
and this number is shown in column 5.
Only $2/54=4$\% of these ellipticals are detected in H$\alpha$, so 
we expect $78-4=74$\% of the undetected population to be cluster members.  This final number is the one
we are most interested in, and it is given in column 6.  For the Scd class, 
this number is much smaller.  We compare our measured fraction of 19\% with the expected
fraction of field Scd galaxies in our sample, 12\% at $I\sim19$, to determine that only 37\% of Scd galaxies are
cluster members.  Since we detect H$\alpha$ in 26\% of these galaxies, we conclude that only 11\% of the
undetected Scd galaxies are cluster members.  In practice, these calculations account for
the magnitude-dependence of the field contamination, and include the weights for the sampling fraction
as a function of luminosity and radius (\S\ref{sec-field}).  Also, in some cases (see the Sab example in
Table~\ref{tab-fieldcor}), the expected fraction of undetected cluster members is less than zero,
reflecting the uncertainty and variation in the field composition.  In this case, we assume that none of the undetected
galaxies belong to the cluster.
\begin{table} 
\begin{center} 
\caption{{\sc \label{tab-fieldcor} Statistical field corrections at $I=19$}} 
\vspace{0.1cm}
{\scriptsize
\begin{tabular}{lccccc} 
\hline\hline
\noalign{\smallskip}
(1) & (2)       & (3)     & (4)             & (5)          & (6)               \cr
Type& $N_{\rm gal}$     &$N_{H\alpha}$ & $f_{\rm field}$ & $f_{\rm cluster}$&$f_{\rm cluster}^{\rm no H\alpha}$\cr
\hline
E  &  54 (27\%)         & 2           & 6             &78              &74\cr 
S0 &  79 (40\%)        & 4           & 8             &80              &75\cr 
Sab&  28 (14\%)        & 7           & 12            &14              &   0\cr 
Scd&  38 (19\%)        & 10          & 12            &37              &11\cr 
\noalign{\hrule}
\end{tabular}
}
\end{center} 
\end{table}
\end{document}